\documentclass[lettersize,journal]{IEEEtran}
\usepackage{amsmath,amsfonts,amssymb}
\usepackage{algorithmic}
\usepackage{algorithm}
\usepackage{array}
\usepackage[caption=false,font=footnotesize,labelfont=rm,textfont=rm]{subfig}
\usepackage{textcomp}
\usepackage{stfloats}
\usepackage{url}
\usepackage{verbatim}
\usepackage{graphicx}
\usepackage{cite}
\usepackage{xcolor}

\begin{document}

\title{Joint Optimization of Flexible Antenna Array Shape and Beamforming for Secure Communication}

\author{Zhen Xu, Gaojie Chen, ~\IEEEmembership{Senior Member, IEEE}, Jing Zhu, ~\IEEEmembership{Member, IEEE}, Weiwei Zhao, Yonghui Li, ~\IEEEmembership{Fellow, IEEE}, Rahim Tafazolli, ~\IEEEmembership{Fellow, IEEE}, and Wei Huang
    \thanks{This work was supported in part by the Fundamental and Interdisciplinary Disciplines Breakthrough Plan of the Ministry of Education of China under Grant JYB2025XDXM406; in part  by the National Natural Science Foundation of China Program under Grant 62501652 and in part by the China Postdoctoral Science Foundation under Grant 2025M773513. (\emph{Corresponding authors: Gaojie Chen.})}
    \thanks{Zhen Xu, Gaojie Chen, Jing Zhu, Wei Huang are with the School of Flexible Electronics (SoFE), Sun Yat-Sen University, Shenzhen, Guangdong 518107, China (e-mail: xuzh253@mail2.sysu.edu.cn, zhuj229@mail.sysu.edu.cn, chengj235@mail.sysu.edu.cn, huangw323@sysu.edu.cn).

    Weiwei Zhao is with the State Key Laboratory of Organic Electronics and Information Displays $\&$ Jiangsu Key Laboratory for Biosensors, Institute of Advanced Materials (IAM), Nanjing University of Posts $\&$ Telecommunications, 9 Wenyuan, Nanjing, 210023, China (e-mail: iamwwzhao@njupt.edu.cn).

    Yonghui Li is with the School of Electrical and Information Engineering, The University of Sydney, Sydney, NSW 2006, Australia (e-mail: yonghui.li@sydney.edu.au).

    Rahim Tafazolli is with 5G and 6G Innovation Centre, Institute for Communication Systems (ICS) of University of Surrey, Guildford, GU2 7XH, UK (e-mail: r.tafazolli@surrey.ac.uk).
    }
    
}
\maketitle

\begin{abstract}
Flexible antenna arrays (FAAs) can physically re-shape their geometry to add new spatial degrees-of-freedom, whereas transmit beamforming adjusts the complex element weights to electronically steer and shape the array’s radiation pattern, thereby significantly improving communication performance. This paper is the first to explore the integration of FAA geometry control and beamforming for physical layer security enhancement, where a base station equipped with an FAA communicates with a legitimate user in the presence of passive eavesdroppers. To safeguard confidential transmissions, we formulate a new secrecy-rate maximization problem that jointly optimizes the transmit beamforming vector and a continuous FAA shape control parameter. Due to the non-convex nature of the problem, an alternating optimization algorithm is developed to decompose the joint design into tractable subproblems, which are solved iteratively to refine both the FAA geometry and beamforming strategy. Simulation results confirm that the proposed joint optimization framework significantly outperforms conventional fixed-shape or beamforming-only schemes, demonstrating the potential of FAA-enabled reconfigurability for secure wireless communications.

\end{abstract}

\begin{IEEEkeywords}
Flexible antenna arrays, physical layer security, secrecy rate, alternating optimization.
\end{IEEEkeywords}

\section{INTRODUCTION}
\IEEEPARstart{T}{HE} rapid evolution of sixth-generation (6G) wireless networks is driven by the growing need for ultra-reliable, low-latency, and high-capacity connectivity to support transformative applications such as autonomous systems, holographic communications, and the Internet of Everything \cite{wang2023road}. These emerging use cases impose stringent requirements on data rates, spectral efficiency, and massive connectivity, thereby pushing current wireless technologies to their operational limits.
Multiple-input multiple-output (MIMO) technology has been widely acknowledged as a key enabler to meet these demands, owing to its capability to exploit spatial multiplexing and diversity \cite{larsson2014massive,wang2019overview,wang2024tutorial}. By utilizing multiple antennas at both the transmitter and receiver, MIMO systems can transmit multiple parallel data streams and combat channel fading, significantly enhancing throughput and link reliability. Consequently, MIMO remains a cornerstone of both existing and future wireless standards.
However, traditional MIMO implementations typically employ fixed-position antenna elements arranged on rigid, non-adaptive platforms. While effective in static or predictable environments, such architectures inherently lack spatial reconfigurability, limiting their ability to adapt to user mobility, dynamic multipath propagation, and spatially selective interference. Moreover, in compact platforms such as wearable devices or unmanned aerial vehicles (UAVs), physical size constraints further restrict the deployment of large antenna arrays and the exploitation of spatial degrees of freedom. These limitations underscore the urgent need for reconfigurable antenna technologies that can introduce additional spatial flexibility and dynamically adapt to the propagation environment, thereby enhancing the efficiency and resilience of MIMO systems \cite{zheng2025electromagnetically,wang2024ai,new2024tutorial}.

Among emerging reconfigurable antenna technologies, fluid antenna (FA), also known as movable antenna (MA), has attracted significant attention in the context of next-generation wireless systems \cite{zhu2024historical}. Unlike traditional rigid arrays, an FA allows its radiating element to move continuously within a predefined spatial region, effectively leveraging position diversity to reshape the wireless channel. This spatial reconfigurability facilitates enhanced signal reception, improved interference avoidance, and higher multiplexing efficiency, especially in scenarios with limited physical space or rapidly varying propagation conditions \cite{wong2020fluid,zhu2023movable,zhu2024fluid}. Prior research on FA systems has largely focused on aspects such as channel modeling, estimation techniques, performance evaluation, and antenna position optimization. Specifically, Wong \textit{et al}. developed a spatial correlation-based channel model tailored for FA systems in \cite{wong2022bruce,wong2021fluid}, which was later complemented by the introduction of a field-response-based channel model in \cite{zhu2023modeling,mei2024movable}. Building upon these models, several channel estimation approaches were proposed, targeting both the spatial correlation framework \cite{skouroumounis2022fluid,xu2023channel} and the field-response setting \cite{ma2023compressed}. Regarding antenna positioning strategies, the literature presents two main paradigms: discrete port selection schemes \cite{chai2022port} and continuous spatial trajectory optimization methods \cite{ma2023mimo,qin2024antenna,wu2023movable,zhang2024efficient}. Furthermore, key performance metrics such as outage probability and diversity order have been analytically studied in recent works \cite{vega2024fluid,new2023fluid,psomas2023diversity}, providing deeper insight into the reliability of FA-assisted systems under different propagation conditions. Moreover, the authors \cite{zhu2024index} introduced fluid antenna index modulation (FA-IM), a transmission scheme integrating index modulation with FA-assisted MIMO for enhanced spectral efficiency and lower hardware costs. In parallel with FA and MA, rotatable antenna (RA) has recently emerged as another promising flexible antenna technology, which dynamically alters the three-dimensional boresight directions of directional antennas to reconfigure the spatial radiation pattern, thereby providing additional degrees of freedom for wireless communication and sensing \cite{zheng2025rotatable,11222668}.

Despite repositioning individual antenna elements, recent research has shifted attention toward flexible antennas, which are fabricated from stretchable, bendable, and lightweight conductive materials deposited on deformable substrates \cite{kirtania2020flexible,alhaddad2021flexible}. Unlike FAs/MAs that rely on mechanical repositioning, flexible antennas achieve spatial adaptability through structural deformation such as bending, twisting, and folding. This capability not only allows the antenna to conform to non-planar or wearable surfaces but also enables the formation of adaptive radiation patterns across different frequencies and incidence angles, thereby enhancing electromagnetic performance in dynamic environments. Building upon this foundation, the concept has evolved into a flexible antenna array (FAA), which extends structural reconfigurability to multi-element arrays \cite{gan2023bendable,al2021flexible,yang2025flexible}. By allowing the entire array to deform through bending or folding, FAAs support tunable inter-element spacing, array curvature, and orientation in real time. These properties enable precise control over beamforming behavior, side lobe levels, and spatial selectivity. Compared with fluid or movable antennas where a single radiating element is moved within a small region to exploit location diversity \cite{zhu2024historical}, a flexible antenna array reshapes the geometry of a multi element aperture through rotation, bending, or folding of a deformable substrate \cite{yang2025flexible}. This structural reconfigurability jointly modifies the array manifold and the orientation of the element patterns, so that the transmitter can tune both the digital beamformer and the physical aperture to match the propagation environment. From a security perspective, these extra spatial degrees of freedom allow the system to concentrate energy toward the legitimate user while steering deep nulls and strongly reduced side lobes toward potential eavesdroppers, which provides additional secrecy gains beyond those offered by fluid or movable antenna based designs. Current research on FAA is primarily concentrated in the antenna domain, investigating their characteristics from multiple perspectives, such as radiation performance and structural adaptability \cite{gan2023bendable,al2021flexible}. In contrast, the authors of \cite{yang2025flexible} applied FAAs to wireless communication networks, where the channel behavior under different array deformation states is modeled and corresponding performance analysis is conducted.

From a materials and device perspective, recent experimental progress indicates that mechanically reconfigurable antenna structures compatible with the proposed FAA models can be realized using state-of-the-art flexible conductors and substrates. In particular, Zhao \cite{zhao20232d} have reported a printed Ti$_3$C$_2$ MXene ultrawideband monopole antenna that preserves impedance matching and radiation gain over a 1.7 4.0 GHz band under repeated bending with radii of only a few centimeters and after more than 1000 bending cycles, while enabling fluent real-time wireless transmission in the bent state \cite{zhao2022flexible}. Building on the same MXene platform, lightweight multi-level superimposed Ti$_3$C$_2$ films have achieved electromagnetic-interference (EMI) shielding effectiveness above 80 dB together with excellent mechanical flexibility \cite{bai2022biocompatible}. Furthermore, MXene-based hydrogels and organohydrogels, such as cellulose/MXene and MXene/poly(acrylic acid) (PAA) networks, have been engineered into biocompatible, stretchable, and compressible strain sensors that remain conductive under large tensile and compressive deformations and long-term cyclic loading \cite{bai2023stretchable,ding2023multifunctional}. Comprehensive reviews on flexible transparent supercapacitors also demonstrate mature fabrication routes for large-area, mechanically robust transparent electrodes and integrated flexible devices \cite{zhao2021flexible}. Collectively, these works show that multi-element flexible panels made of printed MXene conductors and soft substrates are already experimentally feasible, and that the rotations, bendings, and foldings assumed in our FAA model are well aligned with practical materials and manufacturing constraints.

On the other hand, as wireless networks continue to evolve toward open, heterogeneous, and interference-prone environments, physical layer security (PLS) has emerged as a critical design paradigm to safeguard confidential transmissions. Unlike traditional cryptographic techniques that rely on computational complexity, PLS leverages the inherent randomness and spatial characteristics of the wireless medium to ensure secure communication at the signal level \cite{9702524}. Mainstream PLS techniques primarily include artificial noise (AN) \cite{zhou2018artificial}, cooperative jamming \cite{cao2020improving} and secure beamforming \cite{lin2018robust}, which have been extensively studied to impair the eavesdropper’s reception while maintaining reliable communication with legitimate users. Beyond these classical approaches, recent efforts have turned toward reconfigurable antenna technologies, which offer additional spatial degrees of freedom. By dynamically adjusting the antenna geometry or radiation pattern, such systems can further strengthen the legitimate link while simultaneously deteriorating the wiretap channel, thereby improving physical-layer secrecy performance \cite{cheng2024enabling,tang2024secure,hu2024movable,tang2023fluid,ghadi2024physical}. Specifically, \cite{cheng2024enabling} and \cite{tang2024secure} explored strategies to enhance secrecy performance by jointly adjusting the transmit beamforming and the spatial placement of MAs in the presence of an eavesdropper. In addition, when facing multiple eavesdroppers equipped with single antennas, the authors of \cite{hu2024movable} introduced an alternating optimization method based on projected gradient ascent to coordinate the configuration of MAs and beamforming design, thereby strengthening the physical layer security of the system. Wong \textit{et al.} investigated PLS in FA-aided systems under arbitrary fading correlations, deriving closed-form secrecy metrics via copula and quadrature methods, and showed superior performance over conventional schemes \cite{tang2023fluid,ghadi2024physical}.

Compared with single-element FA or conventional fixed position antenna arrays, FAAs introduce a richer set of spatial degrees of freedom (DoF) that can be exploited for secure communication. Specifically, the ability to jointly optimize array shape and beamforming opens up new possibilities for manipulating the spatial characteristics of the wireless channel in favor of the legitimate user while suppressing information leakage to potential eavesdroppers.

To the best of our knowledge, however, using FAA structures in physical layer security, which adds extra spatial DoF and boosts PLS, is still at an early stage. To address this gap, this paper investigates a downlink multiple-input single-output (MISO) wiretap scenario, where a base station equipped with an FAA communicates with a legitimate user in the presence of passive eavesdroppers. We formulate a secrecy rate maximization problem that jointly optimizes the transmit beamforming vector and a continuous FAA shape control parameter. To tackle the resulting non-convex problem, we develop an efficient alternating optimization (AO) algorithm, which iteratively updates the beamforming and array-shape variables. The main contributions of this paper are summarized as follows:

\begin{itemize}
\item
We propose a new joint optimization framework that integrates beamforming design with FAA shape reconfiguration to enhance physical layer security. By jointly adapting the FAA’s geometry and the transmit weight vector, the proposed scheme breaks through the inherent limitations of traditional fixed-position arrays, offering a new degree of spatial control. This dual-domain adaptability significantly enhances the secrecy rate by strengthening the legitimate communication link while simultaneously suppressing the information leakage to eavesdroppers.

\item  
A secrecy-rate maximization problem is formulated that simultaneously optimizes the continuous FAA shape-control parameter and the beamforming vector. Hidden convexity in each subproblem is exposed through variable decoupling, enabling efficient solutions despite the non-convex global objective.

\item
An alternating-optimization algorithm is developed that iteratively refines the beamformer-obtained in closed form as the dominant generalized eigenvector of a Hermitian matrix pair-and the array shape, updated via projected gradient ascent within its feasible interval.  The procedure provably yields a non-decreasing secrecy rate, converges to a first-order stationary point, and requires only $\mathcal{O}(N)$ complexity per iteration.

\item

Simulation results show that the proposed joint optimization framework clearly outperforms fixed‐shape and beamforming‐only schemes. In the single‐eavesdropper case, it offers $0.7\;\mathrm{bps/Hz}$ higher secrecy rate than beamforming‐only, representing a 16\% gain, and $1.9\;\mathrm{bps/Hz}$ higher than shape‐only, representing a 58\% gain. It also maintains a steady $0.7\;\mathrm{bps/Hz}$ advantage when multiple eavesdroppers are present.

\end{itemize}

The remainder of this paper is organized as follows. In Section II, we present the system model and the formulated problem. Section III introduces the alternating optimization algorithm. Section IV provides the numerical results. Finally, we conclude the paper. 

\textit{Notations:} Scalars are in italic; vectors in bold lowercase; matrices in bold uppercase; calligraphic letters denote sets. The Hermitian transpose is $(\cdot)^{\mathrm H}$, magnitude $|\cdot|$, Euclidean norm $\|\cdot\|_2$, positive-part operator $[\cdot]^+$, and real part $\Re\{\cdot\}$. The carrier wavelength is $\lambda$. System parameters: $N_h,N_v$ are the numbers of horizontal and vertical antenna elements; $N=N_hN_v$ is the total element count; $d$ is the inter-element spacing; $\psi$ is the FAA shape-control variable; $\mathbf{r}_n(\psi)\in\mathbb R^{3}$ is the position of the $n$th element; $\mathbf{w}\in\mathbb C^{N}$ is the transmit beamformer; $P_{\max}$ is the power limit. Channel quantities: $\mathbf{h}_B(\psi),\mathbf{h}_E(\psi)\in\mathbb C^{N}$ are the channels to Bob and Eve; $\alpha_{B,l},\alpha_{E,l}$ are the $l$th path gains; $\sigma^{2}$ is the noise variance.

\section{System Model}
We consider a downlink secure communication system as depicted in Fig. 1, where a base station (BS) equipped with a FAA comprising $N = N_h\times N_v$ antenna elements communicates with a single-antenna legitimate receiver, Bob, in the presence of $M$ passive eavesdroppers, denoted by $\{{\rm{Eve}}_i\}_{i=1}^M$, each equipped with a single antenna.
The BS conveys information to Bob while Eves attempt to intercept the transmission. We assume that the transmitter has perfect channel state information (CSI) for both Bob and Eves to enable optimal transmission design. The FAA’s deformation is governed by a single shape‐control parameter \(\psi\), which quantifies the degree of flexibility for each FAA configuration. The base station can reshape the array by tuning \(\psi\) and simultaneously adjust the transmit beamforming vector \(\mathbf{w}\in\mathbb{C}^{N\times1}\) to create channel conditions that favor Bob and disadvantage Eves. We focus on a single‐cell, small‐coverage scenario in which both Bob and Eves lie in the base station’s far‐field region and remain quasi‐static. This setup facilitates a tractable analysis of joint beamforming and array-shape design. The base station employs beamforming to transmit to the legitimate user Bob. Let the complex transmit symbol be \(s\) and the beamforming vector be \(\mathbf{w}=[w_1, w_2, \dots, w_N]^T\in\mathbb{C}^{N\times1}\). Accordingly, the signal emitted by antenna \(n\) is \(w_n\,s\), with all antennas radiating the same waveform coherently. In vector form, the transmitted signal is  
\begin{equation}
    \mathbf{x} \;=\; \mathbf{w}s,
\end{equation}
subject to the total power constraint  
\begin{equation}
\|\mathbf{w}\|^2 \;\le\; P_{\max}.
\end{equation}
On the downlink, \(\mathbf{x}\) propagates over a multipath wireless channel and is received by both Bob and $M$ Eves. 

\subsection{Channel Model}
To facilitate the understanding of this system, we first present the channel modeling under a single-eavesdropper scenario, which will be later extended to the case with multiple eavesdroppers. We adopt a narrowband block-fading multipath model for the BS-to-user channels.  In particular, there are \(L_B\) independent paths between the base station and Bob, and \(L_{E}\) paths between the BS and Eve.  The complex gain of Bob’s \(l\)th path is \(\alpha_{B,l}\), which incorporates both path loss and phase, and its azimuth and elevation angles are \(\phi_{B,l}\) and \(\theta_{B,l}\).  Likewise, Eve’s \(l\)th path has gain \(\alpha_{{E},l}\) and azimuth and elevation angles are \(\phi_{E,l}\) and \(\theta_{E,l}\). The FAA’s shape parameter \(\psi\) affects each path’s array response and antenna gain.  Let the position of the BS’s \(n\)th antenna element be  
\begin{equation}
\mathbf{r}_n(\psi) = \bigl[x_n(\psi),\,y_n(\psi),\,z_n(\psi)\bigr].
\end{equation}

In line with the FAA deformation models presented in \cite{yang2025flexible}, we categorize and model the array shape under three representative configurations: rotatable, bendable, and foldable, as detailed below.

\textit{1) Flexible rotating model:} In this model, the $(n_h,n_v)$th element’s coordinates are given by
\begin{equation}
\left\{
\begin{aligned}
{x}_{n} &= -\frac{2n_h - N_h - 1}{2}d \sin\psi,\quad n_h \in \{1, \cdots, N_h\}, \\
{y}_{n} &= \frac{2n_h - N_h - 1}{2}d \cos\psi,\quad n_h \in \{1, \cdots, N_h\}, \\
{z}_{n} &= \frac{2n_v - N_v - 1}{2}d,\quad n_v \in \{1, \cdots, N_v\},
\end{aligned}
\right.
\end{equation}
where $n_h\in\{1,\dots,N_h\}$ and $n_v\in\{1,\dots,N_v\}$ denote the antenna indices along the horizontal and vertical dimensions, respectively, and $d$ is the uniform inter‐element spacing. $\mathbf{E}_{\text{rot}}=
A_E(\theta,\phi-\psi)\,\mathbf 1_{N_h}\mathbf 1_{N_v}^{\!T}$. Here \(A_E(\theta,\phi)\) denotes the complex‐valued radiation coefficient of an individual antenna element in the spherical direction \((\theta,\phi)\), it characterises the element’s amplitude response, including its intrinsic pattern, polarisation, and the orientation change caused by array rotation, bending, or folding. When the array shape changes, the local element coordinate system also rotates, so the effective azimuth seen by the element becomes \(\phi-\psi_{n_h}\). For an omnidirectional element we simply have \(A_E(\theta,\phi)=1\); for a directional element \(A_E(\theta,\phi)\) is obtained from the power pattern \(G_E(\theta,\phi)\) via \(A_E(\theta,\phi)=\sqrt{G_E(\theta,\phi)}\). By stacking these rows, \(\mathbf{E}(\theta,\phi,\psi)\) compactly captures how every element’s boresight shift \(\psi_{n_h}\) modulates its gain toward an incoming wave from elevation \(\theta\) and azimuth \(\phi\). 

\begin{figure}[!t]
\centering
\includegraphics[width=2.5in]{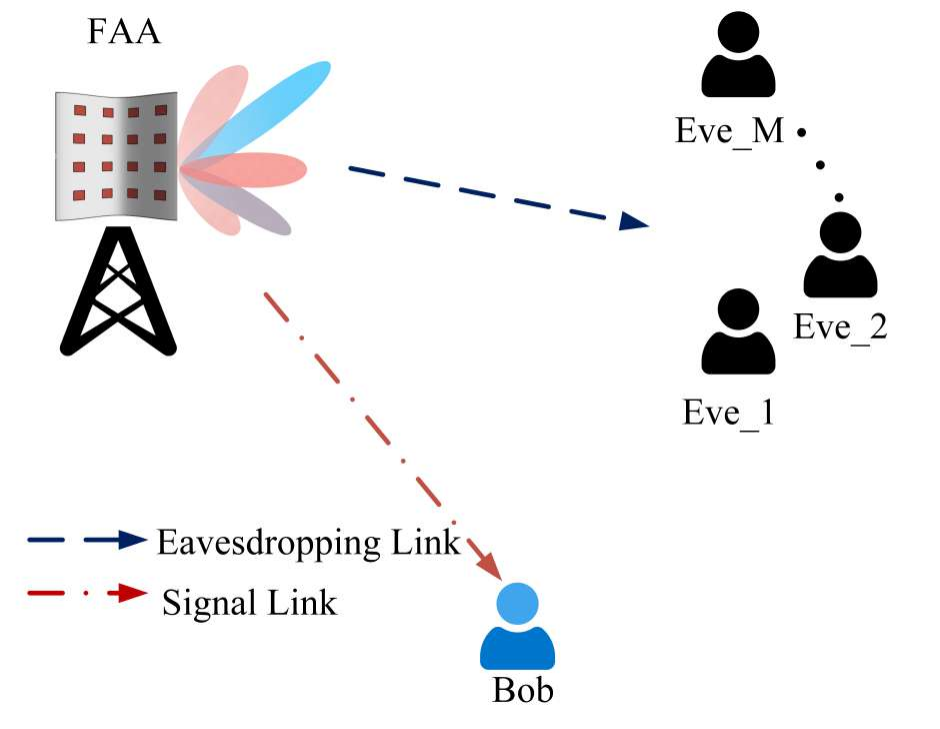}
\caption{A secure communication system assisted by FAA.}
\label{fig_1}
\end{figure}

\textit{2) Flexible bending model:} For the flexible bending model, the coordinates of each antenna in the FAA of $N_v \times N_h$ after the bend can be expressed as
\begin{equation}
\left\{
\begin{aligned}
x_{n} &= R(\cos(\psi_{n_h}) - 1), \quad n_h \in \{1, \cdots, N_h\}, \\
y_{n} &= R\sin(\psi_{n_h}), \quad n_h \in \{1, \cdots, N_h\}, \\
z_{n} &= \frac{2n_v - N_v - 1}{2}d, \quad n_v \in \{1, \cdots, N_v\},
\end{aligned}
\right.
\end{equation}
where $R = \frac{(N_h - 1)d}{2\psi}$ and $\psi_{n_h}=-\psi+2\psi\frac{(n_h-1)}{N_h-1}.$ The radiation pattern for all elements is encapsulated in a pattern matrix 
\(\mathbf{E}_{\text{bend}}(\theta, \phi, \psi) \in \mathbb{C}^{N_h \times N_v}\), which is expressed as
\begin{equation}
\mathbf{E}_{\text{bend}}(\theta, \phi, \psi) \triangleq 
\begin{bmatrix}
A_E(\theta, \phi - \psi_1) & \cdots & A_E(\theta, \phi - \psi_1) \\
A_E(\theta, \phi - \psi_2) & \cdots & A_E(\theta, \phi - \psi_2) \\
\vdots & \ddots & \vdots \\
A_E(\theta, \phi - \psi_{N_h}) & \cdots & A_E(\theta, \phi - \psi_{N_h})
\end{bmatrix}.
\end{equation}

\textit{3) Flexible folding model:} In this model, we have 
\begin{equation}
\left\{
\begin{aligned}
{x}_{n} &= -\left| \frac{2n_h - N_h - 1}{2} \right| d \sin\psi, \quad n_h \in \{1, \cdots, N_h\}, \\
{y}_{n} &= \frac{2n_h - N_h - 1}{2} d \cos\psi, \quad n_h \in \{1, \cdots, N_h\}, \\
{z}_{n} &= \frac{2n_v - N_v - 1}{2} d, \quad n_v \in \{1, \cdots, N_v\},
\end{aligned}
\right.
\end{equation}
and
\begin{equation}
\mathbf{E}_{\text{fold}}=
     \begin{cases}
       A_E(\theta,\phi+\psi), & n_h\!\le\!N_h/2,\\
       A_E(\theta,\phi-\psi), & n_h\!>\!N_h/2.
     \end{cases}
\end{equation}
The array manifold vector \(\mathbf{a}(\psi,\phi,\theta)\in\mathbb{C}^{N\times1}\) collects the per-element phase shifts 
\begin{equation}
\begin{aligned}
&\mathbf{a}(\psi,\phi,\theta) = \\
&\bigl[e^{-j\mathbf{k}(\phi,\theta)\mathbf{r}_1(\psi)},\;
e^{-j\mathbf{k}(\phi,\theta)\mathbf{r}_2(\psi)},\; \dots,\;
e^{-j\mathbf{k}(\phi,\theta)\mathbf{r}_N(\psi)}\bigr]^T,
\end{aligned}
\end{equation}
where \(\lambda\) is the carrier wavelength and  
\(\mathbf{k}(\phi,\theta) = \tfrac{2\pi}{\lambda}[\sin\theta\cos\phi,\;\sin\theta\sin\phi,\;\cos\theta]^T\)  
is the far‐field wavevector.  Because \(\mathbf{r}_n\) depends on \(\psi\), it follows that \(\mathbf{a}(\psi,\phi,\theta)\) also depends on \(\psi\). 

If the antenna channel model assumes an omnidirectional antenna, then $A_E(\theta,\phi)=1$, when it is a directional antenna, 
\begin{equation}
A_E(\theta,\phi)=\sqrt{G_E(\theta,\phi)}, 
\end{equation}
where $G_E(\theta,\phi)$ is the radiation power, according to \cite{10910066}, its expression is 
\begin{equation}
G_E(\theta, \phi) = 
\begin{cases}
G \sin^\kappa \theta \cos^\kappa \phi, & \theta \in \left[0, \frac{\pi}{2} \right],\ \phi \in [0, 2\pi] \\
0, & \text{otherwise}
\end{cases}, 
\end{equation}
where $G=2(\kappa+1)$ is the factor used for normalization. $\kappa$ denotes the pattern‑sharpness factor of each antenna element. In this work, we assume the inter-element spacing along the flexible surface remains sufficient during deformation to avoid severe mutual coupling effects. While deformation may induce minor variations in coupling impedance, we focus on the dominant impact of geometric phase shifts on beamforming and secrecy performance, leaving the complex electromagnetic coupling modeling for future work.

It is worth noting that when the FAA is in a rotatable state, $\psi$ effectively changes the representation of all path angles $\phi_{B,l}$ and $\phi_{E,l}$ in the array coordinate system; when the FAA is bent/folded, different paths have different incident angles on different array elements, but overall $\psi$ can change the composite gain of the array for a specific direction. We assume that there are $L_B$ paths for Bob and $L_E$ paths for Eve, the channel is established by
\begin{equation}
    \mathbf{h_B}(\psi) \;=\; \sqrt{\frac{1}{L_B}} \sum_{l=1}^{L_B} \alpha_{B,l}\;\mathbf{g}( \phi_{B,l})\,\odot\,\mathbf{a}(\theta_{l}, \phi_{l}, \psi),
\end{equation}
\begin{equation}
    \mathbf{h_E}(\psi) \;=\; \sqrt{\frac{1}{L_E}} \sum_{l=1}^{L_E} \alpha_{E,l}\;\mathbf{g}(\phi_{E,l})\,\odot\,\mathbf{a}(\theta_{l}, \phi_{l}, \psi),
\end{equation}
where $\mathbf{g}(\theta,\phi,\psi) \triangleq \mathrm{vec}\left(\mathbf{E}(\theta, \phi, \psi)\right) \in \mathbb{C}^{N_h N_v \times 1}$ comprises the radiation patterns of all elements, representing the column-stacking for three models. $\odot$ denotes the Hadamard product.

\subsection{Problem Formulation}
\textit{1) Single-eavesdropper scenario:} After the signal $\mathbf{x}=\mathbf{w}s$ is transmitted at the transmitter and propagated through the above channel, Bob and Eve receive the signal after superimposing multipath. For the legitimate user Bob, the received signal can be expressed as
\begin{equation}
y_B \;=\; \mathbf{h_B}(\psi)^H \mathbf{w}s \;+\; n_B\,,
\end{equation}
where $\mathbf{h_B}(\psi)^H \mathbf{w} = \sum_{n=1}^N h_{B,n}(\psi)^* w_n$ represents the equivalent complex gain after beamforming through the channel, $n_B$ is the receiving noise, which is modeled as complex additive white Gaussian noise (AWGN), and $n_B\sim \mathcal{CN}(0,\sigma_B^2)$. Since we assume single-user downlink communication and the transmitter performs beamforming on Bob, there is no interference from other users in Bob's received signal. Similarly, Eve's received signal is
\begin{equation}
y_E \;=\; \mathbf{h_E}(\psi)^H \mathbf{w}s \;+\; n_E\,,
\end{equation}
where $\mathbf{h_E}(\psi)^H \mathbf{w}$ is the equivalent complex gain of the signal under Eve's channel, and $n_E\sim \mathcal{CN}(0,\sigma_E^2)$ is the noise received by Eve. For the convenience of analysis, we assume that the noise power of Bob and Eve is the same, that is, $\sigma_B^2=\sigma_E^2=\sigma^2$. From the above signal model, it can be seen that: under the given beamforming $\mathbf{w}$ and array shape $\psi$, the signal-to-noise ratio $\mathrm{SNR}_B = \frac{|\mathbf{h}_B^H \mathbf{w}|^2}{\sigma^2}$ received by Bob, and the signal-to-noise ratio $\mathrm{SNR}_E = \frac{|\mathbf{h}_E^H \mathbf{w}|^2}{\sigma^2}$ received by Eve. When $\mathbf{h}_B$ and $\mathbf{h}_E$ are spatially independent enough, the transmitter can choose $\mathbf{w}$ to face the direction of $\mathbf{h}_B$ and form an antenna null or low gain in the direction of $\mathbf{h}_E$. The addition of flexible arrays further provides a new degree of freedom $\psi$ to achieve this goal: the base station can physically change the shape of the array to make $\mathbf{h}_B(\psi)$ align the beam direction and $\mathbf{h}_E(\psi)$ deviate from the beam, thereby essentially changing the channel gain relationship.

To evaluate the robustness of the proposed scheme against channel uncertainty, we consider the imperfect CSI model\cite{8543651}. Specifically, the estimated channel vector $\hat{\mathbf{h}}_E(\psi)$ for Eve is modeled as
\begin{equation}
  \hat{\mathbf{h}}_E(\psi)
  = \sqrt{1 - \xi^2}\,\mathbf{h}_E(\psi) + \xi\,\mathbf{e}_E,
\end{equation}
where $\mathbf{h}_E(\psi)$ denotes the true channel vector, and $\mathbf{e}_E \sim \mathcal{CN}(\mathbf{0}, \sigma_e^2 \mathbf{I}_N)$ represents the channel estimation error, which is independent of $\mathbf{h}_E(\psi)$. The parameter $\xi \in [0, 1]$ indicates the CSI quality, with $\xi = 0$ corresponding to the perfect-CSI case. In our simulations, the error vector $\mathbf{e}_E$ is normalized such that it has the same average power as $\mathbf{h}_E(\psi)$. Since the estimation error mainly arises from receiver noise or pilot contamination and is independent of the array geometry, we assume $\frac{\mathrm{d} \mathbf{e}_E}{\mathrm{d}\psi} = \mathbf{0}$. Consequently, the gradient used for the optimization with imperfect CSI is given by
\begin{equation}
  \frac{\mathrm{d}\hat{\mathbf{h}}_E(\psi)}{\mathrm{d}\psi}
  = \sqrt{1 - \xi^2}\,\frac{\mathrm{d}\mathbf{h}_E(\psi)}{\mathrm{d}\psi}.
\end{equation}
It is worth noting that only the CSI of Eve is assumed to be imperfect, while the CSI of Bob is perfectly known at the BS. In all imperfect CSI simulations, the BS designs the beamformer and the array shape parameter based on $\hat{\mathbf{h}}_E(\psi)$ and the perfect CSI of Bob, whereas the secrecy rate is evaluated using the true channels.

In this system, assuming that both Bob and Eve use the optimal Gaussian coding for capacity implementation, the channel capacity of Bob $C_B$ and the channel capacity of Eve $C_E$ are
\begin{equation}
C_B = \log_2\!\Big(1 + \frac{|\mathbf{h}_B(\psi)^H \mathbf{w}|^2}{\sigma^2}\Big),
\end{equation}
\begin{equation}
C_E = \log_2\!\Big(1 + \frac{|\mathbf{h}_E(\psi)^H \mathbf{w}|^2}{\sigma^2}\Big)\,.
\end{equation}
Then the maximum secrecy rate that can be achieved is the positive part of the difference between the two, that is
\begin{equation} \label{eqars}
\begin{aligned}
&R_s(\mathbf{w},\psi) = \Big[\,C_B \;-\; C_E\,\Big]^+ \\
&=\left[\,\log_2\!\Big(1 + \frac{|\mathbf{h}_B^H \mathbf{w}|^2}{\sigma^2}\Big) \;-\; \log_2\!\Big(1 + \frac{|\mathbf{h}_E^H \mathbf{w}|^2}{\sigma^2}\Big)\,\right]^+\,,
\end{aligned}
\end{equation}
where $[x]^+ \triangleq \max(x,0)$ ensures that when the eavesdropping channel capacity exceeds the legitimate channel capacity, the secrecy rate is recorded as 0. The above expression clearly shows that improving the secrecy rate requires increasing the signal-to-noise ratio at Bob and reducing the signal-to-noise ratio at Eve. The secure transmission optimization problem for single eavesdropper system is thus formulated as
\begin{equation}\label{P1}
\begin{aligned}\text{(P1)}: \quad  \max_{\mathbf{w},\,\psi} \quad & R_s(\mathbf{w},\psi) \\
\text{s.t.}\quad & \|\mathbf{w}\|_2^2 \;\le\; P_{\max}, \\[3pt]
& \psi_{\min} \;\le\; \psi \;\le\; \psi_{\max}\,. 
\end{aligned} 
\end{equation}
 \noindent 

It is worth emphasizing that the FAA deformation is governed by a single real-valued shape-control variable $\psi$, which is constrained to lie in the compact interval $[\psi_{\min},\psi_{\max}]$ in (21). This interval models practical implementation limits such as the maximum admissible rotation angle, bending radius, or folding stroke that avoid plastic deformation or fatigue of the flexible substrate and actuators. Moreover, $\psi$ is assumed to be updated on a quasi-static timescale, e.g., when user positions or large-scale channel statistics change, while remaining constant over many symbol intervals. As a result, the mechanical actuation and control loop only need to track a slowly varying scalar parameter instead of per-element positions, so that the required response speed and sensing resolution for “real-time’’ FAA reconfiguration are compatible with existing soft-actuator and flexible-antenna prototypes.

\textit{2) Multiple eavesdropper scenario:}
We consider the case with multiple eavesdroppers, where Alice needs to transmit confidential information to Bob, but there are $M$ eavesdroppers, denoted by $\{\text{Eve}_i\}_{i=1}^M$. Bob and each of $M$ Eves are equipped with a single and fixed-position antenna. In the colluding case, the \(M\) single-antenna eavesdroppers fully cooperate, they exchange their baseband samples and apply joint decoding.  Stacking the individual channel vectors as
\(
\mathbf H_{E}(\psi)
    \triangleq
    \bigl[\mathbf h_{E,1}(\psi),\dots,\mathbf h_{E,M}(\psi)\bigr]
    \in\mathbb C^{N\times M},
\)
the cooperating eavesdroppers behave like an \(M\)-antenna receiver that
performs optimal maximum-ratio combining (MRC).  The resulting
post-combining signal-to-noise ratio (SNR) is
\begin{equation}
    \mathrm{SNR}_{E}^{\mathrm{colluding}}
        \;=\;
        \frac{\bigl\lVert \mathbf H_{E}(\psi)^{H}\mathbf w\bigr\rVert^{2}}{\sigma^{2}}
        \;=\;
        \frac{\displaystyle\sum_{i=1}^{M}
              \bigl|\mathbf h_{E,i}(\psi)^{H}\mathbf w\bigr|^{2}}{\sigma^{2}}.
    \label{eq:SNR_E_coop}
\end{equation}
Accordingly, the eavesdroppers’ channel capacity is
\begin{equation}
    C_{E}^{\mathrm{colluding}}
        \;=\;
        \log_{2}\!\Bigl(1+\mathrm{SNR}_{E}^{\mathrm{colluding}}\Bigr).
    \label{eq:CE_coop}
\end{equation}
Finally, substituting \eqref{eq:CE_coop} together with Bob’s capacity
\(C_B\) yields the achievable secrecy rate under colluding
eavesdropping:
\begin{IEEEeqnarray}{rCl}\label{eq:Rs_coop_multiEve}
R_{s}^{\mathrm{colluding}}(\mathbf w,\psi)
 &=& \Bigl[
       \log_{2}\!\Bigl(
         1+\frac{\bigl|\mathbf h_{B}^{H}(\psi)\mathbf w\bigr|^{2}}{\sigma^{2}}
       \Bigr)
       \nonumber\\
 &&\quad -\;
       \log_{2}\!\Bigl(
         1+\frac{\displaystyle\sum_{i=1}^{M}
                \bigl|\mathbf h_{E,i}^{H}(\psi)\mathbf w\bigr|^{2}}
                {\sigma^{2}}
       \Bigr)
     \Bigr]^{+}.
\end{IEEEeqnarray}
Replacing the single-Eve secrecy rate \(R_s(\mathbf w,\psi)\) in~\eqref{eqars}
with its cooperative counterpart \(R_s^{\text{colluding}}(\mathbf w,\psi)\)
in~\eqref{eq:Rs_coop_multiEve} leads to the following optimization problem
\begin{equation}\label{P2}
\begin{aligned}
\text{(P2)}:\quad
 \max_{\mathbf w,\;\psi}\;&\;
     R_{s}^{\mathrm{colluding}}(\mathbf w,\psi) \\[4pt]
\text{s.t.}\quad
 & \|\mathbf w\|_{2}^{2}\;\le\;P_{\max},\\[2pt]
 & \psi_{\min}\;\le\;\psi\;\le\;\psi_{\max},
\end{aligned}
\end{equation}
which inherits the non-convexity of (P1) and incorporates the additional bound $\psi_{\min} \le \psi \le \psi_{\max}$ that captures the limited deformation range of the shape angle; the objective now depends on all eavesdroppers through the aggregated SNR term~\eqref{eq:Rs_coop_multiEve}, making the joint design of $\mathbf w$ and $\psi$ even more challenging.

 \section{Proposed Algorithm}
   In this section, we propose an alternating optimization algorithm to solve (P1) and (P2). In this section, we first address the optimization problem for the single eavesdropper scenario and later extend the analysis to the multiple eavesdropper scenario.

\subsection{Secrecy rate optimization for the single eavesdropper}
\subsubsection{Transmit Beamformer Optimization}
\label{subsec:beam_opt}

In this subsection we derive a {\it closed-form} solution for the optimal
beam-forming vector~$\mathbf w$ when the array‐shape parameter~$\psi$ is kept
fixed.  Throughout the derivation we write
\[
\mathbf h_B=\mathbf h_B(\psi),\qquad
\mathbf h_E=\mathbf h_E(\psi),
\]
and omit the explicit dependence on~$\psi$ for brevity. With \eqref{eqars} the secrecy rate reads\footnote{Because each AO
iteration will be shown to increase the secrecy rate monotonically, the
positive‐part operator~$[\cdot]^+$ can be dropped without altering the
maximiser.}
\begin{equation}\label{eq:Rs_w_only}
  \max_{\|\mathbf w\|_2^2\le P_{\max}}
  R_s(\mathbf w)
      =\max_{\|\mathbf w\|_2^2\le P_{\max}}
      \log_2\!\left(\frac{1+\gamma_B(\mathbf w)}{1+\gamma_E(\mathbf w)}\right),
\end{equation}
where $\gamma_B(\mathbf w) \triangleq
      \tfrac{|\mathbf h_B^{\mathrm H}\mathbf w|^2}{\sigma^{2}}$ and
      $\gamma_E(\mathbf w) \triangleq
      \tfrac{|\mathbf h_E^{\mathrm H}\mathbf w|^2}{\sigma^{2}}$.
Because the logarithm is monotone, \eqref{eq:Rs_w_only} is equivalent to
\begin{equation}\label{eq:fractional_form}
  \max_{\|\mathbf w\|_2^2\le P_{\max}}
  \;F(\mathbf w)\triangleq
  \frac{1+\gamma_B(\mathbf w)}{1+\gamma_E(\mathbf w)}.
\end{equation}
Introduce the power‐normalised vector
$\mathbf u\triangleq \mathbf w/\sqrt{P_{\max}}$ 
(\,$\|\mathbf u\|_2\le 1$\,) and the constant
$c\triangleq\sigma^{2}/P_{\max}$.  Then
\[
\gamma_B(\mathbf w)=\frac{|\mathbf h_B^{\mathrm H}\mathbf u|^2}{c},
\quad
\gamma_E(\mathbf w)=\frac{|\mathbf h_E^{\mathrm H}\mathbf u|^2}{c},
\]
and the objective in \eqref{eq:fractional_form} becomes
\begin{equation}\label{eq:rayleigh_numer_denom}
  F(\mathbf u)=
  \frac{c+|\mathbf h_B^{\mathrm H}\mathbf u|^2}
       {c+|\mathbf h_E^{\mathrm H}\mathbf u|^2}
  =\frac{\mathbf u^{\mathrm H}\!
          \bigl(\mathbf h_B\mathbf h_B^{\mathrm H}+c\mathbf I\bigr)
          \mathbf u}
         {\mathbf u^{\mathrm H}\!
          \bigl(\mathbf h_E\mathbf h_E^{\mathrm H}+c\mathbf I\bigr)
          \mathbf u},
  \qquad
  \|\mathbf u\|_2\le 1,
\end{equation}
which denotes a generalised Rayleigh
quotient.  By the classical result for such quotients, its maximum is
$\lambda_{\max}\bigl(\mathbf B^{-1}\mathbf A\bigr)$, attained at the dominant
generalised eigenvector of the pair
$(\mathbf A,\mathbf B)$, where $\mathbf A\triangleq\mathbf h_B\mathbf h_B^{\mathrm H}+c\mathbf I$ and $\mathbf B\triangleq\mathbf h_E\mathbf h_E^{\mathrm H}+c\mathbf I.$

Consequently, for any fixed array shape~$\psi$, the secrecy-rate-maximising
beamforming vector is
\begin{equation}\label{prop:w_star}
  \mathbf w^\star(\psi) =
  \sqrt{P_{\max}}\;
  \frac{\mathbf v_{\max}\!\bigl(\mathbf B^{-1}\mathbf A\bigr)}
       {\bigl\|\mathbf v_{\max}\!\bigl(\mathbf B^{-1}\mathbf A\bigr)\bigr\|_2},
\end{equation}
where $\mathbf v_{\max}(\cdot)$ denotes the dominant eigenvector and
$c=\sigma^{2}/P_{\max}$.

\paragraph*{Computational complexity}
~\eqref{prop:w_star} requires the principal eigenvector of an
$N\times N$ Hermitian matrix.  Using a power or Lanczos method with
$N_\mathrm{it}$ iterations the cost is $\mathcal O(N^2N_\mathrm{it})$,
which is negligible compared with the channel estimation overhead in a
typical TDD block when $N<128$.

\subsubsection{Array–Shape Parameter Optimization}
\label{subsec:shape_opt}

Let the transmit beamformer obtained in \eqref{prop:w_star} be denoted henceforth by $\mathbf{w}_{\mathrm{fixed}}$.  With the beamformer held constant, the secrecy‑rate maximization problem (P1) simplifies to a one‑dimensional search over the array‑shape parameter.

\begin{equation}
\begin{aligned}
\text{(P3)}:\quad
\max_{\psi} \quad & R_s(\mathbf{w}_{\text{fixed}},\psi) \\
\text{s.t.}\quad & \psi_{\min} \;\le\; \psi \;\le\; \psi_{\max}.
\end{aligned}
\end{equation}
Since $\log_2(\cdot)$ is monotonically increasing, and from the previous content we know that $[\cdot]^+$ can be ignored, maximizing the secrecy rate is equivalent to maximizing the ratio
\begin{equation}
    \max_{\psi} F(\psi) \triangleq \frac{1 + \frac{|\mathbf{h}_B(\psi)^H \mathbf{w}_{\text{fixed}}|^2}{\sigma^2}}{1 + \frac{|\mathbf{h}_E(\psi)^H \mathbf{w}_{\text{fixed}}|^2}{\sigma^2}} = \frac{\sigma^2 + |\mathbf{h}_B(\psi)^H \mathbf{w}_{\text{fixed}}|^2}{\sigma^2 + |\mathbf{h}_E(\psi)^H \mathbf{w}_{\text{fixed}}|^2}.
\end{equation}
Assume $S_B(\psi) = |\mathbf{h}_B(\psi)^H \mathbf{w}_{\text{fixed}}|^2$, $S_E(\psi) = |\mathbf{h}_E(\psi)^H \mathbf{w}_{\text{fixed}}|^2$, then the objective is to maximize
\begin{equation}
    F(\psi) = \frac{\sigma^2 + S_B(\psi)}{\sigma^2 + S_E(\psi)} ,
\end{equation}
which is a nonlinear univariate optimization problem with interval constraints and is nonconvex. We will use the gradient ascent method to solve it, this method will converge to a local maximum, but the quality of the solution can be improved by using multiple starting points. Using the derivative rule of the quotient
\begin{equation}\label{30}
     \frac{dF(\psi)}{d\psi} = \frac{\frac{dS_B(\psi)}{d\psi}(\sigma^2 + S_E(\psi)) - (\sigma^2 + S_B(\psi))\frac{dS_E(\psi)}{d\psi}}{(\sigma^2 + S_E(\psi))^2}.
\end{equation}

Next, we use the universal channel $\mathbf{h}(\psi)$ to represent $\mathbf{h}_B(\psi)$ or $\mathbf{h}_E(\psi)$ and find its derivative $\frac{dS(\psi)}{d\psi}$
\begin{equation}
\begin{aligned}
     &\frac{dS(\psi)}{d\psi} = \frac{df(\psi)}{d\psi}f(\psi)^* + f(\psi)\frac{df(\psi)^*}{d\psi} = 2 \text{Re}\left\{ \frac{df(\psi)}{d\psi}f(\psi)^* \right\} \\
     &=2 \text{Re}\left\{ \left(\left(\frac{d\mathbf{h}(\psi)}{d\psi}\right)^H \mathbf{w}\right) (\mathbf{w}^H \mathbf{h}(\psi)) \right\}
     \end{aligned}
\end{equation}
where $f(\psi) = \mathbf{h}(\psi)^H \mathbf{w}$, so $S(\psi) = |\mathbf{h}(\psi)^H \mathbf{w}|^2 = (\mathbf{h}(\psi)^H \mathbf{w}) (\mathbf{w}^H \mathbf{h}(\psi))=f(\psi)f(\psi)^*$. By using the derivation rule, we get (32). The next step is to find $\frac{d\mathbf{h}(\psi)}{d\psi}$ from (14) and (15). Suppose $\mathbf{c}_l(\psi) = \mathbf{g}( \phi_{l}, \theta_l, \psi)\,\odot\,\mathbf{a}(\theta_{l}, \phi_{l}, \psi)$, then
\begin{equation}
    \frac{d\mathbf{h}(\psi)}{d\psi} = \sqrt{\frac{1}{L}} \sum_{l=1}^{L} \alpha_{l}\;\frac{d\mathbf{c}_l(\psi)}{d\psi}.
\end{equation}
Using the product rule of the Hadamard product
\begin{equation}
    \frac{d\mathbf{c}_l(\psi)}{d\psi} = \frac{d\mathbf{g}_l(\psi)}{d\psi} \odot \mathbf{a}_l(\psi) + \mathbf{g}_l(\psi) \odot \frac{d\mathbf{a}_l(\psi)}{d\psi},
\end{equation} 
where $\mathbf{g}_l(\psi) = \mathbf{g}(\phi_l, \theta_l, \psi)$, $\mathbf{a}_l(\psi) = \mathbf{a}(\theta_l, \phi_l, \psi)$. The next step is to find the derivatives of $\mathbf{g}_l(\psi)$ and $\mathbf{a}_l(\psi)$. The $n$th element of $\mathbf{g}_l(\psi)$ is $g_{l,n}(\psi) = A_E(\theta_l, \phi'_{l,n}(\psi))$, where $\phi'_{l,n}(\psi)$ is the effective azimuth of path $l$ at antenna $n$.
\begin{equation}
\begin{aligned}
    A_E(\theta_l, \phi'_{l,n}(\psi)) &= \sqrt{G \sin^\kappa \theta_l \cos^\kappa (\phi'_{l,n}(\psi))} \\&= (\sqrt{G \sin^\kappa \theta_l}) (\cos (\phi'_{l,n}(\psi)))^{\kappa/2}
    \end{aligned}.
\end{equation}
Assume $K_{\theta_l} = \sqrt{G \sin^\kappa \theta_l}$, then $g_{l,n}(\psi) = K_{\theta_l} (\cos (\phi'_{l,n}(\psi)))^{\kappa/2}$. Its derivative is $\frac{dg_{l,n}(\psi)}{d\psi} = K_{\theta_l} \cdot \frac{d}{d\psi} (\cos (\phi'_{l,n}(\psi)))^{\kappa/2}$. Using the chain rule,we have
\begin{align}
\frac{d g_{l,n}(\psi)}{d\psi}
  &=
  K_{\theta_l}\,\frac{\kappa}{2}\,
  \bigl[\cos(\phi'_{l,n}(\psi))\bigr]^{\frac{\kappa}{2}-1}
  \nonumber\\[2pt]
  &\quad \times
  \bigl(-\sin(\phi'_{l,n}(\psi))\bigr)
  \frac{d\phi'_{l,n}(\psi)}{d\psi}.
\label{eq:dg_dpsi}
\end{align}

This derivative is valid when $\cos(\phi'_{l,n}(\psi)) > 0$, which corresponds to the angular region where the antenna element has non-zero gain according to the power pattern defined in (13).

\paragraph{Rotatable Model}
For all elements $n$ in the array, the effective azimuth is $\phi'_{l,n}(\psi) = \phi_l - \psi$. Hence, its derivative with respect to $\psi$ is $\frac{d\phi'_{l,n}(\psi)}{d\psi} = -1$. Substituting this into the general derivative expression for $g_{l,n}(\psi)$, we get
\begin{equation}
    \frac{dg_{l,n}(\psi)}{d\psi} = K_{\theta_l} \frac{\kappa}{2} (\cos(\phi_l-\psi))^{\frac{\kappa}{2}-1}  \sin(\phi_l-\psi).
\end{equation}
Note that the two negative signs from the chain rule and $\frac{d\phi'_{l,n}}{d\psi}$ cancel each other out. Since this expression is independent of the element index $n$, all elements of the vector $\frac{d\mathbf{g}_l(\psi)}{d\psi}$ are identical.

\paragraph{Bendable Model}
The effective azimuth for an element in the $n_h$th row is $\phi'_{l,n}(\psi) = \phi_l - \psi_{n_h}(\psi)$, where the local rotation angle is $\psi_{n_h}(\psi) = \left(2\frac{n_h-1}{N_h-1}-1\right)\psi \triangleq K'_{n_h}\psi$. Therefore, the derivative is $\frac{d\phi'_{l,n}(\psi)}{d\psi} = -K'_{n_h}$. Substituting this into the derivative expression for $g_{l,n}(\psi)$ yields
\begin{equation}
    \frac{dg_{l,n}(\psi)}{d\psi} = K'_{n_h} K_{\theta_l}\frac{\kappa}{2} (\cos(\phi_l - K'_{n_h}\psi))^{\frac{\kappa}{2}-1}\sin(\phi_l - K'_{n_h}\psi).
\end{equation}
The negative sign from the chain rule and the negative sign from $-K'_{n_h}$ cancel out. The elements of the vector $\frac{d\mathbf{g}_l(\psi)}{d\psi}$ are the same for all antennas with the same horizontal index $n_h$, but differ across rows with different $n_h$.

\paragraph{Foldable Model}
The derivative of the gain vector $\frac{d\mathbf{g}_l(\psi)}{d\psi}$ depends on which half of the array the element belongs to.

\begin{itemize}
    \item For the first half ($n_h \le N_h/2$): The effective azimuth is $\phi'_{l,n}(\psi) = \phi_l + \psi$, so $\frac{d\phi'_{l,n}(\psi)}{d\psi} = 1$. The derivative of the gain element is
    \begin{equation}
        \frac{dg_{l,n}(\psi)}{d\psi} = - K_{\theta_l}\frac{\kappa}{2} (\cos(\phi_l + \psi))^{\frac{\kappa}{2}-1}\sin(\phi_l + \psi).
    \end{equation}
    Here, the negative sign comes from the derivative of the cosine term.

    \item For the second half ($n_h > N_h/2$): The effective azimuth is $\phi'_{l,n}(\psi) = \phi_l - \psi$, so $\frac{d\phi'_{l,n}(\psi)}{d\psi} = -1$. The derivative of the gain element is
    \begin{equation}
        \frac{dg_{l,n}(\psi)}{d\psi} = K_{\theta_l}\frac{\kappa}{2} (\cos(\phi_l - \psi))^{\frac{\kappa}{2}-1}\sin(\phi_l - \psi).
    \end{equation}
    In this case, the negative sign from the cosine's derivative and the negative sign from $\frac{d\phi'_{l,n}}{d\psi}$ cancel each other. For an odd‑indexed array ($N_h$ odd), the central column remains undeformed, so its local azimuth does not vary with $\psi$, i.e., $\tfrac{d\phi'_{l,n}}{d\psi}=0$ for $n_h=\tfrac{N_h+1}{2}$.

\end{itemize}
Thus, the elements of the vector $\frac{d\mathbf{g}_l(\psi)}{d\psi}$ are divided into two distinct groups based on the horizontal index $n_h$.

Derivative of $\mathbf{a}_l(\psi)$:

The $n$th element of $\mathbf{a}_l(\psi)$ is $a_{l,n}(\psi) = e^{-j\mathbf{k}(\phi_l,\theta_l)\cdot\mathbf{r}_n(\psi)}$.
Let $\mathbf{k}_l = \mathbf{k}(\phi_l,\theta_l)$, then
\begin{equation}
    \frac{da_{l,n}(\psi)}{d\psi} = a_{l,n}(\psi) \left(-j\mathbf{k}_l\frac{d\mathbf{r}_n(\psi)}{d\psi}\right),
\end{equation}
where $\frac{d\mathbf{r}_n(\psi)}{d\psi} = \left[\frac{dx_n(\psi)}{d\psi}, \frac{dy_n(\psi)}{d\psi}, \frac{dz_n(\psi)}{d\psi}\right]^T$. The derivation element $\frac{d\mathbf{r}_n(\psi)}{d\psi}$ of each model is calculated as follows:
\setcounter{paragraph}{0} 
\paragraph{Rotatable Model}
$x_n(\psi) = -C_{n_h} d \sin\psi$, $\frac{dx_n}{d\psi} = -C_{n_h} d \cos\psi$ 
$y_n(\psi) = C_{n_h} d \cos\psi$, $\frac{dy_n}{d\psi} = -C_{n_h} d \sin\psi$ 
$z_n(\psi) = C_{n_v} d$, $\frac{dz_n}{d\psi} = 0$ 
Where $C_{n_h} = \frac{2n_h - N_h - 1}{2}$. 
therefore: 
\begin{equation}
    \mathbf{k}_l\frac{d\mathbf{r}_n(\psi)}{d\psi} = -\frac{2\pi d C_{n_h}}{\lambda} \sin\theta_l \cos(\phi_l-\psi)
\end{equation}
\paragraph{Bendable Model}
Radius of curvature $R(\psi) = \frac{(N_h - 1)d}{2\psi}$, $\frac{dR}{d\psi} = -\frac{R(\psi)}{\psi}$. 
$\psi_{n_h}(\psi) = K'_{n_h}\psi$, $\frac{d\psi_{n_h}}{d\psi} = K'_{n_h}$ ($K'_{n_h} = 2\frac{n_h-1}{N_h-1}-1$). 
$x_n(\psi) = R(\psi)(\cos(\psi_{n_h}(\psi)) - 1)$ 
$\frac{dx_n}{d\psi} = -\frac{R}{\psi}(\cos(K'_{n_h}\psi)-1) - R K'_{n_h} \sin(K'_{n_h}\psi)$ 
$y_n(\psi) = R(\psi)\sin(\psi_{n_h}(\psi))$ 
$\frac{dy_n}{d\psi} = -\frac{R}{\psi}\sin(K'_{n_h}\psi) + R K'_{n_h} \cos(K'_{n_h}\psi)$ 
$\frac{dz_n}{d\psi} = 0$ 
therefore: 
\begin{equation}
    \mathbf{k}_l\frac{d\mathbf{r}_n(\psi)}{d\psi} = \frac{2\pi}{\lambda} \left( \sin\theta_l\cos\phi_l \frac{dx_n}{d\psi} + \sin\theta_l\sin\phi_l \frac{dy_n}{d\psi} \right)
\end{equation}
 \paragraph{Foldable Model}
$C_{n_h,abs} = -\left| \frac{2n_h - N_h - 1}{2} \right|$, $C_{n_h,lin} = \frac{2n_h - N_h - 1}{2}$. 
$x_n(\psi) = C_{n_h,abs} d \sin\psi$, $\frac{dx_n}{d\psi} = C_{n_h,abs} d \cos\psi$ 
$y_n(\psi) = C_{n_h,lin} d \cos\psi$, $\frac{dy_n}{d\psi} = -C_{n_h,lin} d \sin\psi$ 
$\frac{dz_n}{d\psi} = 0$ 
therefore: 
\begin{equation}
\begin{aligned}
    \mathbf{k}_l \cdot \frac{d\mathbf{r}_n(\psi)}{d\psi} &= \frac{2\pi d}{\lambda} \left( (\sin\theta_l\cos\phi_l)(C_{n_h,abs}\cos\psi) \right. \\ 
    &\quad \left. + (\sin\theta_l\sin\phi_l)(-C_{n_h,lin}\sin\psi) \right). 
\end{aligned}
\end{equation}
Finally, the closed form gradients of $F(\psi)$ for the three models are collected in
\eqref{eq:grad_rot}–\eqref{eq:grad_fold}. When the optimal position of $\psi$ is found by the PGA method based on the AdaGrad algorithm, the corresponding $\psi$ in the $(v + 1)$ iteration is first given by the following formula
\begin{equation}
(\psi)^{\nu+1} = (\psi)^{\nu} + \delta_{\text{adat}}\frac{dF(\psi)}{d\psi},
\end{equation}
where $\delta_{\nu}$ is the step size at iteration $\nu$. Then, the result is projected back onto the feasible interval $[\psi_{\min}, \psi_{\max}]$
\begin{align}
\psi^{\nu+1} &=
  \text{proj}_{[\psi_{\min}, \psi_{\max}]}\!\bigl(\tilde{\psi}^{\nu+1}\bigr) \notag\\
             &=
  \max\!\bigl(\psi_{\min},\, \min(\tilde{\psi}^{\nu+1},\,\psi_{\max})\bigr).
\end{align}
Since $\psi$ is a one-dimensional continuous variable and $F(\psi)$ admits an analytical gradient form as given in (31)–(46), the PGA can rapidly converge to a first-order stationary point with a complexity of $\mathcal{O}(LN N_{\text{PGA}})$. Compared with dense grid search, PGA requires a lower computational cost for the same accuracy and can be easily interleaved with the beamforming update in (27). To enhance robustness, we initialize $\psi$ with multiple starting points and update them using the same \emph{Adam} step-size rule, which yields a stable monotonically increasing curve in practice. To avoid numerical overshooting, we truncate $dF/d\psi$ and project $\psi$ onto $[\psi_{\min},\psi_{\max}]$ after each step.

 \begin{figure*}[!b]
	\textsc{\centering
		\hrulefill
}
\begingroup\small
\begin{align}\label{eq:grad_rot}
\frac{dF_{\mathrm{rot}}(\psi)}{d\psi}
&=
\frac{
  2\,\Re\!\Bigl\{%
      \bigl(\sum_{n}w_n^{*}h_{B,n}^{\mathrm{rot}}\bigr)^{*}
      \sum_{n}w_n^{*}\dot{h}_{B,n}^{\mathrm{rot}}
  \Bigr\}\!
  \Bigl(\sigma^{2}+|\sum_{n}w_n^{*}h_{E,n}^{\mathrm{rot}}|^{2}\Bigr)
 -
  2\,\Re\!\Bigl\{%
      \bigl(\sum_{n}w_n^{*}h_{E,n}^{\mathrm{rot}}\bigr)^{*}
      \sum_{n}w_n^{*}\dot{h}_{E,n}^{\mathrm{rot}}
  \Bigr\}\!
  \Bigl(\sigma^{2}+|\sum_{n}w_n^{*}h_{B,n}^{\mathrm{rot}}|^{2}\Bigr)}
 {%
  \Bigl(\sigma^{2}+|\sum_{n}w_n^{*}h_{E,n}^{\mathrm{rot}}|^{2}\Bigr)^{2}},
\\[4pt]
h_{u,n}^{\mathrm{rot}}(\psi)
&=\notag
\frac{1}{\sqrt{L_{u}}}
\sum_{l=1}^{L_{u}}
\alpha_{u,l}\,
\sqrt{G}\,
\sin^{\kappa/2}\!\theta_{u,l}\,
\cos^{\kappa/2}\!\bigl(\phi_{u,l}-\psi\bigr)
\\[2pt]
&\quad{}\times
\exp\!\Bigl\{-j\frac{2\pi d}{\lambda}
  \bigl[
     -C_{n_h}\sin\psi\,\sin\theta_{u,l}\cos\phi_{u,l}
     +C_{n_h}\cos\psi\,\sin\theta_{u,l}\sin\phi_{u,l}
     +C_{n_v}\cos\theta_{u,l}
  \bigr]\Bigr\},
\\[6pt]
\dot{h}_{u,n}^{\mathrm{rot}}(\psi)
&=\notag
\frac{1}{\sqrt{L_{u}}}
\sum_{l=1}^{L_{u}}
\alpha_{u,l}\,
\sqrt{G}\,
\sin^{\kappa/2}\!\theta_{u,l}\,
\cos^{\kappa/2-1}\!\bigl(\phi_{u,l}-\psi\bigr)
\\[2pt]
&\quad{}\times\notag
\Bigl[
     \tfrac{\kappa}{2}\sin\!\bigl(\phi_{u,l}-\psi\bigr)
   + j\frac{2\pi d C_{n_h}}{\lambda}\,
     \sin\theta_{u,l}\cos\!\bigl(\phi_{u,l}-\psi\bigr)
\Bigr]
\\[2pt]
&\quad{}\times
\exp\!\Bigl\{-j\frac{2\pi d}{\lambda}
  \bigl[
     -C_{n_h}\sin\psi\,\sin\theta_{u,l}\cos\phi_{u,l}
     +C_{n_h}\cos\psi\,\sin\theta_{u,l}\sin\phi_{u,l}
     +C_{n_v}\cos\theta_{u,l}
  \bigr]\Bigr\}.
\end{align}
\endgroup

\begingroup\small
\begin{align}
\frac{dF_{\mathrm{bend}}(\psi)}{d\psi}
&=
\frac{
  2\,\Re\!\Bigl\{%
      \bigl(\sum_{n}w_n^{*}h_{B,n}^{\mathrm{bend}}\bigr)^{*}
      \sum_{n}w_n^{*}\dot{h}_{B,n}^{\mathrm{bend}}
  \Bigr\}\!
  \Bigl(\sigma^{2}+|\sum_{n}w_n^{*}h_{E,n}^{\mathrm{bend}}|^{2}\Bigr)
 -
  2\,\Re\!\Bigl\{%
      \bigl(\sum_{n}w_n^{*}h_{E,n}^{\mathrm{bend}}\bigr)^{*}
      \sum_{n}w_n^{*}\dot{h}_{E,n}^{\mathrm{bend}}
  \Bigr\}\!
  \Bigl(\sigma^{2}+|\sum_{n}w_n^{*}h_{B,n}^{\mathrm{bend}}|^{2}\Bigr)}
 {%
  \Bigl(\sigma^{2}+|\sum_{n}w_n^{*}h_{E,n}^{\mathrm{bend}}|^{2}\Bigr)^{2}},
\\[4pt]
h_{u,n}^{\mathrm{bend}}(\psi)
&=\notag
\frac{1}{\sqrt{L_{u}}}
\sum_{l=1}^{L_{u}}
\alpha_{u,l}\,
\sqrt{G}\,
\sin^{\kappa/2}\!\theta_{u,l}\,
\cos^{\kappa/2}\!\bigl(\phi_{u,l}-K'_{n_h}\psi\bigr)
\\[2pt]
&\quad{}\times
\exp\!\Bigl\{-j\frac{2\pi}{\lambda}
  \bigl[
      \sin\theta_{u,l}\cos\phi_{u,l}\,x_{n}(\psi)
     +\sin\theta_{u,l}\sin\phi_{u,l}\,y_{n}(\psi)
     +\cos\theta_{u,l}\,C_{n_v}d
  \bigr]\Bigr\},
\\[6pt]
\dot{h}_{u,n}^{\mathrm{bend}}(\psi)
&=\notag
\frac{1}{\sqrt{L_{u}}}
\sum_{l=1}^{L_{u}}
\alpha_{u,l}\,
\sqrt{G}\,
\sin^{\kappa/2}\!\theta_{u,l}\,
\cos^{\kappa/2-1}\!\bigl(\phi_{u,l}-K'_{n_h}\psi\bigr)
\\[2pt]
&\quad{}\times\notag
\Bigl[
   \tfrac{\kappa}{2}K'_{n_h}\sin\!\bigl(\phi_{u,l}-K'_{n_h}\psi\bigr)
 + j\frac{2\pi}{\lambda}
   \bigl(
      \sin\theta_{u,l}\cos\phi_{u,l}\,\tfrac{dx_{n}}{d\psi}
     +\sin\theta_{u,l}\sin\phi_{u,l}\,\tfrac{dy_{n}}{d\psi}
   \bigr)
\Bigr]
\\[2pt]
&\quad{}\times
\exp\!\Bigl\{-j\frac{2\pi}{\lambda}
  \bigl[
      \sin\theta_{u,l}\cos\phi_{u,l}\,x_{n}(\psi)
     +\sin\theta_{u,l}\sin\phi_{u,l}\,y_{n}(\psi)
     +\cos\theta_{u,l}\,C_{n_v}d
  \bigr]\Bigr\}.
\end{align}
\endgroup

\begingroup\small
\begin{align}
\frac{dF_{\mathrm{fold}}(\psi)}{d\psi}
&=
\frac{
  2\,\Re\!\Bigl\{%
      \bigl(\sum_{n}w_n^{*}h_{B,n}^{\mathrm{fold}}\bigr)^{*}
      \sum_{n}w_n^{*}\dot{h}_{B,n}^{\mathrm{fold}}
  \Bigr\}\!
  \Bigl(\sigma^{2}+|\sum_{n}w_n^{*}h_{E,n}^{\mathrm{fold}}|^{2}\Bigr)
 -
  2\,\Re\!\Bigl\{%
      \bigl(\sum_{n}w_n^{*}h_{E,n}^{\mathrm{fold}}\bigr)^{*}
      \sum_{n}w_n^{*}\dot{h}_{E,n}^{\mathrm{fold}}
  \Bigr\}\!
  \Bigl(\sigma^{2}+|\sum_{n}w_n^{*}h_{B,n}^{\mathrm{fold}}|^{2}\Bigr)}
 {%
  \Bigl(\sigma^{2}+|\sum_{n}w_n^{*}h_{E,n}^{\mathrm{fold}}|^{2}\Bigr)^{2}},
\\[4pt]
h_{u,n}^{\mathrm{fold}}(\psi)
&=\notag
\frac{1}{\sqrt{L_{u}}}
\sum_{l=1}^{L_{u}}
\alpha_{u,l}\,
\sqrt{G}\,
\sin^{\kappa/2}\!\theta_{u,l}\,
\cos^{\kappa/2}\!\bigl(\phi_{u,l}-s_{n_h}\psi\bigr)
\\[2pt]
&\quad{}\times
\exp\!\Bigl\{-j\frac{2\pi d}{\lambda}
  \bigl[
     -|C_{n_h}|\sin\psi\,\sin\theta_{u,l}\cos\phi_{u,l}
     +C_{n_h}\cos\psi\,\sin\theta_{u,l}\sin\phi_{u,l}
     +C_{n_v}\cos\theta_{u,l}
  \bigr]\Bigr\},
\\[6pt]
\dot{h}_{u,n}^{\mathrm{fold}}(\psi)\label{eq:grad_fold}
&=\notag
\frac{1}{\sqrt{L_{u}}}
\sum_{l=1}^{L_{u}}
\alpha_{u,l}\,
\sqrt{G}\,
\sin^{\kappa/2}\!\theta_{u,l}\,
\cos^{\kappa/2-1}\!\bigl(\phi_{u,l}-s_{n_h}\psi\bigr)
\\[2pt]
&\quad{}\times\notag
\Bigl[
  -\tfrac{s_{n_h}\kappa}{2}\sin\!\bigl(\phi_{u,l}-s_{n_h}\psi\bigr)
 + j\frac{2\pi d}{\lambda}\,
   \bigl(
      s_{n_h}|C_{n_h}|\cos\psi\,\sin\theta_{u,l}\cos\phi_{u,l}
     -C_{n_h}\sin\psi\,\sin\theta_{u,l}\sin\phi_{u,l}
   \bigr)
\Bigr]
\\[2pt]
&\quad{}\times
\exp\!\Bigl\{-j\frac{2\pi d}{\lambda}
  \bigl[
     -|C_{n_h}|\sin\psi\,\sin\theta_{u,l}\cos\phi_{u,l}
     +C_{n_h}\cos\psi\,\sin\theta_{u,l}\sin\phi_{u,l}
     +C_{n_v}\cos\theta_{u,l}
  \bigr]\Bigr\}.
\end{align}
\endgroup

\end{figure*}
\subsubsection{Overall Algorithm}

\begin{algorithm}[!htbp]

\caption{Alternating Optimization for Joint Beamforming--Shape Design Solving (P1)}
\label{alg:ao_faa}

\begin{algorithmic}[1]   

\STATE \textbf{Input:} channels $\{\mathbf h_B(\psi),\mathbf h_E(\psi)\}$, power $P_{\max}$, noise $\sigma^{2}$, shape bounds $[\psi_{\min},\psi_{\max}]$, tolerances $\epsilon_{\mathrm{AO}},\epsilon_{\mathrm{PGA}}$, iteration limits $K_{\mathrm{AO}},K_{\mathrm{PGA}}$.
\STATE \textbf{Initialise:} set $\psi^{(0)}\leftarrow(\psi_{\min}{+}\psi_{\max})/2$; choose $\mathbf w^{(0)}$ (e.g.\ maximum‐ratio towards $\mathbf h_B(\psi^{(0)})$), scale to $\|\mathbf w^{(0)}\|_2^{2}=P_{\max}$; compute $R_s^{(0)}$; set $k\leftarrow0$.

\WHILE{$k<K_{\mathrm{AO}}$}

    \STATE $\mathbf A\leftarrow\mathbf h_B(\psi^{(k)})\mathbf h_B^{\mathrm H}(\psi^{(k)})+\tfrac{\sigma^{2}}{P_{\max}}\mathbf I$,\;
           $\mathbf B\leftarrow\mathbf h_E(\psi^{(k)})\mathbf h_E^{\mathrm H}(\psi^{(k)})+\tfrac{\sigma^{2}}{P_{\max}}\mathbf I$.
    \STATE $\mathbf v_{\max}\leftarrow$ dominant eigenvector of $\mathbf B^{-1}\mathbf A$;\;
           $\mathbf w^{(k+1)}\leftarrow\sqrt{P_{\max}}\mathbf v_{\max}/\|\mathbf v_{\max}\|_2$.

    \STATE $\psi_{(0)}\leftarrow\psi^{(k)}$,\; $t\leftarrow0$.
    \REPEAT
        \STATE Evaluate 
              $F\!\bigl(\psi_{(t)}\bigr)=\dfrac{\sigma^{2}+|\mathbf h_B^{\mathrm H}\mathbf w^{(k+1)}|^{2}}
                                            {\sigma^{2}+|\mathbf h_E^{\mathrm H}\mathbf w^{(k+1)}|^{2}}$
              and its slope $\mathrm d F/\mathrm d\psi$ using~\eqref{30}.
        \STATE $\tilde\psi\leftarrow\psi_{(t)}+\delta\,(\mathrm d F/\mathrm d\psi)$ \textit{(AdaGrad stepsize $\delta$)}.
        \STATE $\psi_{(t+1)}\leftarrow\min\{\psi_{\max},\max\{\psi_{\min},\tilde\psi\}\}$ \textit{(projection)}.
        \STATE $t\leftarrow t+1$.
    \UNTIL{$t=K_{\mathrm{PGA}}$ \textbf{or} $|\psi_{(t)}-\psi_{(t-1)}|<\epsilon_{\mathrm{PGA}}$}
    \STATE $\psi^{(k+1)}\leftarrow\psi_{(t)}$.

    \STATE $R_s^{(k+1)}\leftarrow\log_{2}\!\Bigl(1+\tfrac{|\mathbf h_B^{\mathrm H}(\psi^{(k+1)})\mathbf w^{(k+1)}|^{2}}{\sigma^{2}}\Bigr)
                       -\log_{2}\!\Bigl(1+\tfrac{|\mathbf h_E^{\mathrm H}(\psi^{(k+1)})\mathbf w^{(k+1)}|^{2}}{\sigma^{2}}\Bigr)$.

    \IF{$|R_s^{(k+1)}-R_s^{(k)}|/R_s^{(k+1)}<\epsilon_{\mathrm{AO}}$}
        \STATE \textbf{break}
    \ENDIF
    \STATE $k\leftarrow k+1$
\ENDWHILE

\STATE \textbf{Output:} beamformer $\mathbf w^{(k)}$, shape parameter $\psi^{(k)}$, secrecy rate $R_s^{(k)}$.
\end{algorithmic}

\end{algorithm}

The overall algorithm for solving (P1) is presented in Algorithm 1. We first initialize the transmit beamformer $\mathbf{w}^{(0)}$, the FAA shape-control parameter $\psi^{(0)}$, and compute the initial secrecy rate $R_s^{(0)}$. Then, $\mathbf{w}$ and $\psi$ are optimized in an alternating manner. Specifically, in each iteration of the AO process, we first update $\mathbf{w}$ by solving the problem detailed in Section~\ref{subsec:beam_opt} (resulting in the solution \eqref{prop:w_star}) with $\psi$ fixed. Subsequently, we update $\psi$ by solving the problem in Section~\ref{subsec:shape_opt} using Projected Gradient Ascent (PGA) with $\mathbf{w}$ fixed. 
As the objective value of problem (P1), the secrecy rate $R_s$, is monotonically non-decreasing after each sub-problem optimization and is upper-bounded, the proposed alternating optimization based approach is guaranteed to converge to a stationary point, based on a predefined accuracy $\epsilon$ or a maximum number of iterations.

Besides, the complexities of the sub-problem for optimizing $\mathbf{w}$ and the sub-problem for optimizing $\psi$ within each AO iteration are in the order of $\mathcal{O}(N^2 N_{\text{it,eig}})$ and $\mathcal{O}(L N N_{\text{PGA}})$, respectively, where $N_{\text{it,eig}}$ denotes the number of iterations for the generalized eigenvector computation (e.g., power iteration), $L = \max(L_B, L_E)$ is the maximum number of channel paths, and $N_{\text{PGA}}$ denotes the number of iterations for the PGA method. Overall, if the beamformer optimization step is dominant, the complexity of Algorithm 1 is in the order of $\mathcal{O}(K_{\text{AO}} N^2 N_{\text{it,eig}})$, where $K_{\text{AO}}$ is the total number of AO iterations. 

\subsection{Secrecy rate optimization for multiple eavesdroppers}

Problem~(P2) differs from (P1) only in that the $M$ single-antenna
eavesdroppers fully cooperate, which replaces the rank-one term
$\mathbf h_{E}\mathbf h_{E}^{\mathrm H}$ by the rank-$M$ term
$\mathbf H_{E}\mathbf H_{E}^{\mathrm H}$, where
$
  \mathbf H_{E}(\psi)\!
  \triangleq
  \bigl[\mathbf h_{E,1}(\psi),\dots,\mathbf h_{E,M}(\psi)\bigr]
  \in\mathbb C^{N\times M}.
$

\subsubsection{Beamformer update for fixed shape}
For a given $\psi$, define
\begin{align}
\mathbf A &\triangleq
  \mathbf h_{B}(\psi)\mathbf h_{B}^{\mathrm H}(\psi)
  +\frac{\sigma^{2}}{P_{\max}}\mathbf I, \nonumber \\[2pt]
\mathbf B &\triangleq
  \mathbf H_{E}(\psi)\mathbf H_{E}^{\mathrm H}(\psi)
  +\frac{\sigma^{2}}{P_{\max}}\mathbf I.
\label{eq:A_B_P2}
\end{align}

Then the secrecy-rate objective is a generalised Rayleigh quotient whose
maximiser is
\begin{equation}\label{eq:w_star_P2_clean}
  \mathbf w^{\star}(\psi)
  \;=\;
  \sqrt{P_{\max}}\,
  \frac{\,\mathbf v_{\max}\!\bigl(\mathbf B^{-1}\mathbf A\bigr)}
       {\bigl\|\mathbf v_{\max}\!\bigl(\mathbf B^{-1}\mathbf A\bigr)\bigr\|_{2}} ,
\end{equation}
where $\mathbf v_{\max}(\cdot)$ denotes the dominant eigenvector, exactly
analogous to~\eqref{prop:w_star}.
\begin{figure*}[!t]    
  \centering
  \includegraphics[width=\textwidth]{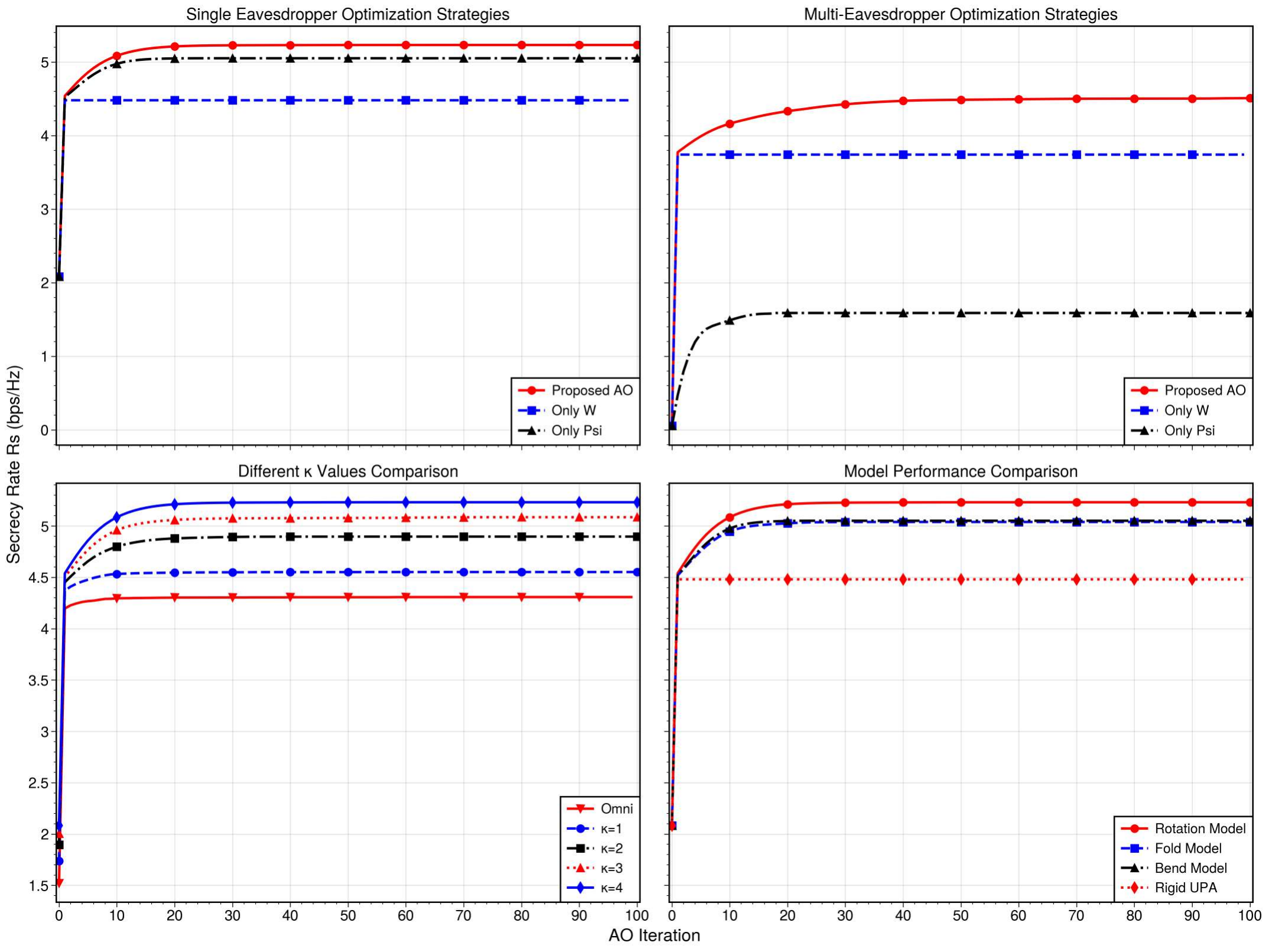}   
  \caption{Convergence of the proposed joint shape–beamforming AO scheme and resulting secrecy‑rate performance across varying eavesdropper scenarios, sharpness factor, and array‑flexibility models.}
  \label{fig_1}
\end{figure*}
\subsubsection{Shape update for fixed beamformer}
With $\mathbf w$ fixed, the objective becomes
\begin{equation}\label{eq:Fpsi_P2_final}
  F(\psi)
  \;=\;
  \frac{\sigma^{2}+|\mathbf h_{B}^{\mathrm H}(\psi)\mathbf w|^{2}}
       {\sigma^{2}+\|\mathbf H_{E}^{\mathrm H}(\psi)\mathbf w\|_{2}^{2}}
  \;\triangleq\;
  \frac{\sigma^{2}+S_{B}(\psi)}{\sigma^{2}+S_{E}(\psi)} .
\end{equation}
The gradient required for the projected-gradient-ascent (PGA) step is
\begin{equation}\label{eq:dF_dpsi_P2_final}
  \frac{\mathrm dF}{\mathrm d\psi}
  \;=\;
  \frac{\bigl(\sigma^{2}+S_{E}\bigr)\dot S_{B}
        -\bigl(\sigma^{2}+S_{B}\bigr)\dot S_{E}}
       {(\sigma^{2}+S_{E})^{2}},
\end{equation}
with
\begin{align}
\dot S_{B} &=
  2\,\operatorname{Re}\!\Bigl\{(\dot{\mathbf h}_{B}^{\mathrm H}\mathbf w)
      (\mathbf w^{\mathrm H}\mathbf h_{B})\Bigr\}, \nonumber \\[2pt]
\dot S_{E} &=
  2\,\operatorname{Re}\!\Biggl\{\sum_{i=1}^{M}
      (\dot{\mathbf h}_{E,i}^{\mathrm H}\mathbf w)
      (\mathbf w^{\mathrm H}\mathbf h_{E,i})\Biggr\}.
\label{eq:dSdpsi_coop}
\end{align}

where $\dot{\mathbf h}\triangleq\frac{\mathrm d\mathbf h}{\mathrm d\psi}$
is obtained exactly as in Sec.~\ref{subsec:shape_opt}.

\subsubsection{Complexity and convergence}
Each AO sweep requires
$\mathcal O\!\bigl(N^{2}N_{\mathrm{eig}}+MN\,N_{\mathrm{PGA}}\bigr)$
floating-point operations, i.e.\ only a factor~$M$ more than in (P1).
As both sub-steps monotonically increase
$R_{s}^{\mathrm{colluding}}$ and that rate is upper-bounded, convergence to a
stationary point is guaranteed.

\section{Numerical Results}
In this section, numerical results are provided to verify the performance of the proposed alternating optimization algorithm for joint FAA shape and beamforming optimization. We consider a MISO system where a base station (BS) equipped with an $N_v \times N_h$ FAA communicates with a legitimate user Bob in the presence of one or more eavesdroppers. Unless otherwise specified, the system parameters are set as follows. The carrier frequency is $f_c = 28$ GHz, corresponding to a wavelength $\lambda = 0.0107$ m. The FAA consists of $N_v=3$ vertical and $N_h=3$ horizontal elements, resulting in $N=N_v N_h = 9$ total elements, with an inter-element spacing of $d = \lambda/2$. The maximum transmit power at the BS is $P_{\max} = 0$ dBm. Bob is assumed to be 50 m from the BS, while in the single-Eve scenario Eve is 80 m away; the specific distances in multi‑eve scenarios will be given where relevant.
 The channel between the BS and each user (Bob/Eve) is modeled with $L=10$ multipath components. The antenna element radiation pattern sharpness factor is $\kappa = 4$. The reference channel power gain at 1 m is $g_0 = -40$ dB, and the path loss exponent is $\alpha = 2.8$. The additive white Gaussian noise (AWGN) power at both Bob and Eve is $\sigma^2 = -92$ dBm. The initial FAA shape-control parameter is set to $\psi^{(0)} = 0^\circ$. The proposed AO algorithm is run for $K_{\text{AO}} = 100$ iterations. The optimization of $\psi$ within each AO step uses an Adam optimizer with a learning rate $\delta_T=0.01$ for $K_{AO} = 100$ iterations or until the change in $\psi$ is less than $10^{-3}$ degrees. All angles are expressed in radians in the analytical derivations; degrees are used in the text/figures for readability and converted to radians in simulations. The presented results are averaged over $N_{\text{MC}} = 1000$ Monte Carlo simulations. The adopted settings of simulation parameters are provided in Table I, unless specified otherwise.
\label{subsec:sol_P2}
\begin{table}
\centering
\caption{Simulation Parameters}
\resizebox{\columnwidth}{!}{%
\begin{tabular}{|c|p{3.8cm}|c|}
\hline
\textbf{Parameter} & \textbf{Description} & \textbf{Value} \\ \hline
$L$ & Number of channel paths & 10 \\ \hline
$\kappa$ & Antenna element radiation pattern sharpness factor & 4 \\ \hline
$\lambda$ & Carrier wavelength & 0.0107 m \\ \hline
$g_0$ & Average channel power gain at reference distance & $-40$ dB \\ \hline
$\alpha$ & Exponent of path loss & 2.8 \\ \hline
$\sigma^2$ & Noise power & $-92$ dBm \\ \hline
$d_{ar}$ & Distance of BS–Bob & 50 m \\ \hline
$d_{ab}$ & Distance of BS–Eve & 80 m \\ \hline
$\delta_T$ & Adam learning rate when optimizing $\psi$ & 0.01 \\ \hline
$K_{AO}$ & Maximum number of AO iterations & 100 \\ \hline
$\tau_T$ & Optimization threshold comparison for $\psi$ & $10^{-3}$ \\ \hline
$\beta_1$ & Adam first‑moment decay rate & 0.9 \\ \hline
$\beta_2$ & Adam second‑moment decay rate & 0.999 \\ \hline
$N_{\text{MC}}$ & Number of Monte Carlo trials & 1000 \\ \hline
\end{tabular}
}
\end{table}
\subsection{Convergence analysis}
Fig. 2 depicts the convergence behaviour and secrecy rate advantages of the proposed joint AO algorithm under other representative schemes. Only $W$ scheme means that only the beamforming vector is optimized, Only $\psi$ scheme means only optimizing $\psi$. In the single‑eavesdropper scenario (upper‑left subfigure) the proposed AO scheme converges within roughly 20~iterations and attains a steady secrecy rate of $5.2\,\mathrm{bps/Hz}$. Compared with the Only~$W$ baseline of $4.5\,\mathrm{bps/Hz}$, which constitutes an absolute gain of $0.7\,\mathrm{bps/Hz}$, corresponding to an approximate $16\%$ relative improvement. The advantage over the Only~$\Psi$ scheme ($3.3\,\mathrm{bps/Hz}$) reaches $1.9\,\mathrm{bps/Hz}$, increased by about 58 percent. When multiple eavesdroppers are present (upper‑right subfigure) the secrecy rate naturally decreases, yet the proposed AO scheme still plateaus at $4.6\,\mathrm{bps/Hz}$, maintaining a margin of $ 0.7\,\mathrm{bps/Hz}$ over Only~$W$ scheme and well above $170\%$ relative to Only~$\Psi$ scheme. In addition to the asymptotic complexity analysis in Section~III, Fig.~2 shows that the proposed AO algorithm converges in fewer than 20 outer iterations for all considered scenarios, irrespective of the FAA model and the number of eavesdroppers. Each outer iteration is dominated by a principal generalized-eigenvector computation of an $N \times N$ Hermitian matrix, complemented by a scalar projected-gradient update of the shape parameter $\psi$. Using standard iterative eigensolvers, this yields an $\mathcal{O}(N^2)$ per-iteration cost, while the $\psi$-update scales only linearly with $N$. Therefore, the overall runtime is comparable to that of conventional secure beamforming algorithms implemented at base stations, which confirms the empirical practicality of the proposed joint shape–beamforming design.

The lower‑left subfigure explores the impact of the element–pattern sharpness factor $\kappa$, both omnidirectional and directional antenna patterns are considered. The legend entry Omni represents omnidirectional antenna pattern and serves as a baseline of $4.3\,\mathrm{bps/Hz}$.  Directional elements are then examined with $\kappa\in\{1,2,3,4\}$ in the pattern model $G_E(\theta,\phi)=2(\kappa+1)\sin^{\kappa}\!\theta\,\cos^{\kappa}\!\phi$.  As $\kappa$ increases from 1 to 4, the secrecy rate rises monotonically, reaching $4.9\,\mathrm{bps/Hz}$ at $\kappa=4$.  This constitutes an absolute gain of $0.6\,\mathrm{bps/Hz}$ over the Omni baseline, corresponding to a relative improvement of approximately $14\%$.  Even the change from $\kappa=1$ to $\kappa=4$ delivers an additional $0.7\,\mathrm{bps/Hz}$ about $16\%$ improvment, highlighting that sharper element patterns and array‑shape agility reinforce each other to enhance secrecy performance across the full range of considered directivities. For a more detailed study of $\kappa$, please refer to the content later on.

Finally, the lower-right subfigure compares three practical shape models with a rigid uniform planar array (UPA). The rotation model, which employs a rotatable array capable of rotating in the horizontal plane, achieves $5.1\,\mathrm{bps/Hz}$ versus $4.5\,\mathrm{bps/Hz}$ for the rigid UPA, yielding a $0.6\,\mathrm{bps/Hz}$ gain, corresponding to an approximate $13\%$ improvement. The fold and bend models follow closely at $4.9\,\mathrm{bps/Hz}$, each sustaining a roughly $9\%$ edge over the rigid array. It is worth noting that in Fig.~2 we adopt a compact $3\times3$ array mainly for visualization purposes, so that the deformation shapes of the three FAA models can be clearly displayed within a single panel. For this odd number of columns, the bend and fold geometries may lead to identical element locations, which in turn produces overlapping secrecy rate curves. This overlap is thus a small array special case rather than an intrinsic equivalence between the two models. In Fig.~4, the total number of antennas $N = N_hN_v$ is increased from $N=9$ to $N=49$, the bend and fold configurations yield distinct optimal shapes and non-overlapping secrecy rate performance.

These observations jointly verify that (i) the proposed AO algorithm exhibits rapid convergence, (ii) simultaneous optimisation of beamforming and array shape is essential in both single and multi‑eavesdropper scenarios, and (iii) introducing realistic shape flexibility consistently enhances secrecy capacity across diverse propagation environments.
\begin{figure}[!t]
\centering
\includegraphics[width=3.2in]{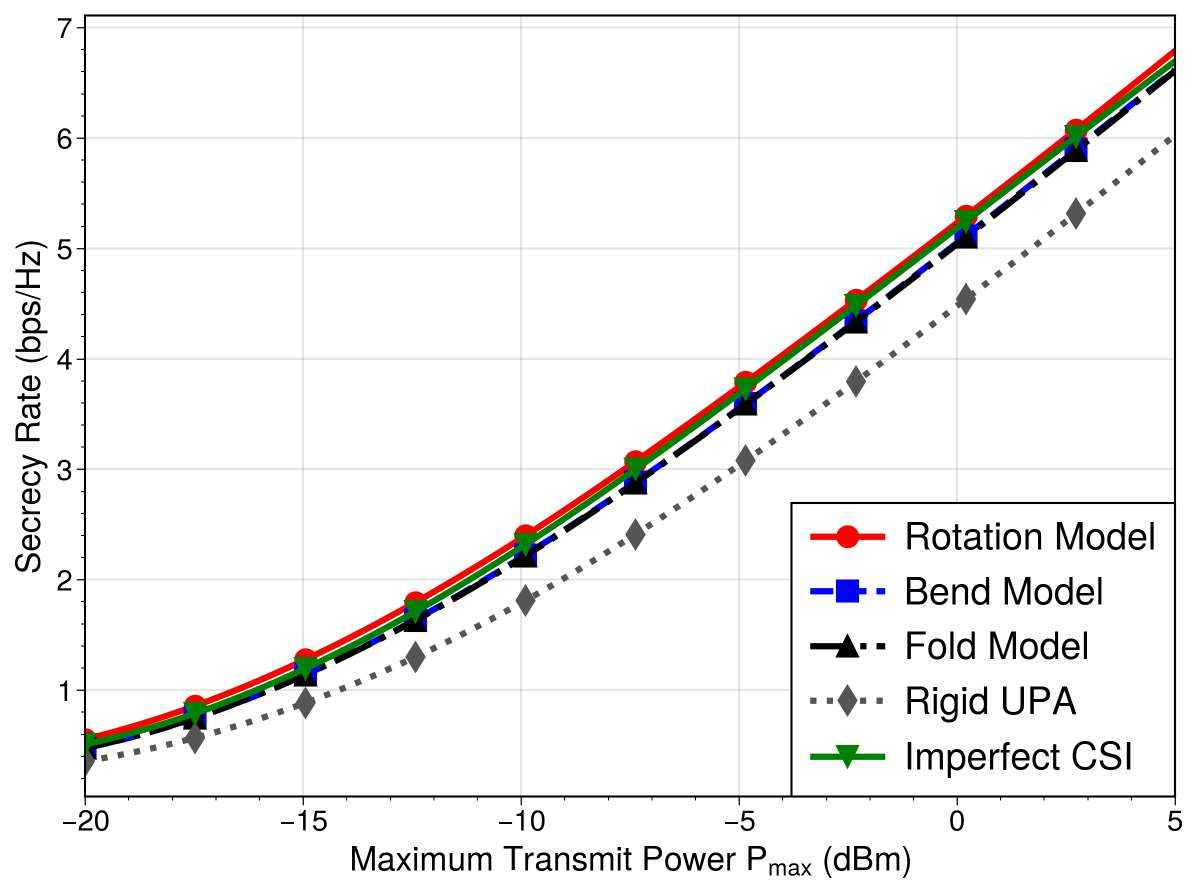}
\caption{Impact of the maximum transmit power on secrecy rate for different configurations.}
\label{fig_2}
\end{figure}

\subsection{Impact of the maximum transmit power}
Fig.~\ref{fig_2} illustrates the achievable secrecy rate $R_s$ versus the maximum transmit power $P_{\max}$ for the rigid planar array and the three flexible‑array models introduced in Section~II. Across the entire range $P_{\max}\!\in[-20,10]\;\mathrm{dBm}$ the rotation, bend, and fold schemes strictly outperform the rigid benchmark, while all curves rise monotonically in accordance with~\eqref{eqars} because both the desired and eavesdropping SNRs scale linearly with the available power yet their difference enters the logarithmic objective. At the practical operating point $P_{\max}=0\;\mathrm{dBm}$ the rotation model attains $R_s\!\approx\!5.5\;\mathrm{bps/Hz}$, offering a $14\%$ gain over the rigid array; the bend and fold schemes both deliver $5.1\;\mathrm{bps/Hz}$—corresponding to $7\%$ improvements, respectively. In Fig.~3, we set $\xi = 0.1$ to emulate a moderate CSI error level. To further illustrate the impact of CSI quality, Fig.~7 plots the secrecy rate as a function of $\xi$ for different array configurations. Under imperfect CSI situation, the performance degrades only marginally relative to the perfect-CSI case, and the schemes still exhibit strong robustness. The reason why the bend and fold curves overlap has been explained previously, so it is omitted here. This performance hierarchy persists throughout the power sweep, reflecting the degrees of freedom each deformation provides: rotation steers the boresight of all elements, whereas bending and folding reshape only subsets of elements and thus face tighter geometric constraints. 

It is worth noting that once $P_{\max}$ becomes sufficiently large, the secrecy‑rate curves will begin to saturate. Under the high‑SNR approximation of~\eqref{eqars},
\[
  R_s \;\xrightarrow{P_{\max}\to\infty}\;
  \log_{2}\!\Bigl(\|\mathbf{h}_{B}\|^{2}\big/\|\mathbf{h}_{E}\|^{2}\Bigr),
\]
which no longer depends on the transmit power. Consequently, in the very high power situation all schemes asymptotically flatten into horizontal lines.

\begin{figure}[!t]
\centering
\includegraphics[width=3.2in]{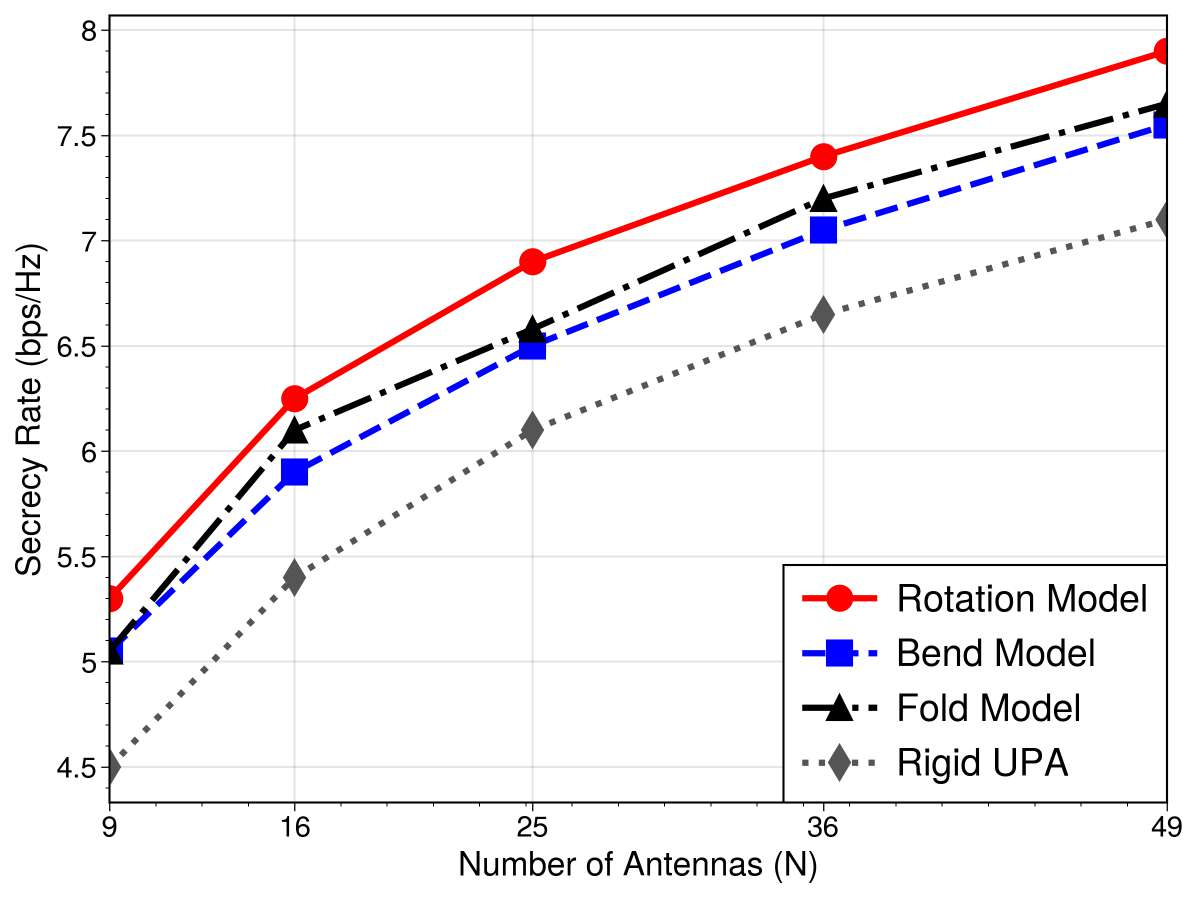}
\caption{Impact of the number of antennas on secrecy rate for different configurations.}
\label{fig_3}
\end{figure}

\subsection{Impact of the array size}
Fig.~\ref{fig_3} plots the achievable secrecy rate versus the total number of elements $N=N_h\times N_v$ for the single Eve scheme with $P_{\max}=0$\,dBm and element sharpness factor $\kappa=4$.  Across all four models: rigid UPA, rotation, bend, and fold, the secrecy rate increases monotonically as the antenna number increases, because a larger array furnishes additional spatial degrees of freedom that sharpen the main‑lobe towards Bob while carving deeper spatial nulls towards Eve.  The rotation‑enabled FAA enjoys the steepest growth: by steering the entire panel it can simultaneously preserve Bob in the boresight and sweep the Eve into sidelobe minima, so its advantage over the rigid benchmark widens with $N$.  The bend and fold models also exploit the extra DoFs, but their gains saturate earlier owing to the partial deformation constraint that only the edge columns are re‑orientable when $N_h=3$, as discussed in detail above. Nevertheless they maintain a consistent margin above the rigid UPA throughout the tested range.  These results verify that shape reconfigurability and antenna number increasing act synergistically rather than redundantly in enhancing physical layer security.

According to the angle mappings in (10) and (14), when $N_h$ is odd both the bending and folding geometries possess an immovable central column, because at $n_h=(N_h+1)/2$ their tilt functions satisfy $\psi^{\text{bend}}_{n_h}= \psi^{\text{fold}}_{n_h}=0$, the corresponding elements keep the original boresight for any shaping parameter~$\psi$. This stationary spine preserves a full gain contribution in the desired direction and, thanks to the statistical symmetry of the remaining columns, renders the aggregate array factors of the two shapes almost identical; consequently their secrecy rate curves almost overlap for the $3\times3$, $5\times5$, $7\times7$… arrays. When $N_h$ is even the spine disappear.  Bending then imposes only a small linear tilt $\pm\psi/(N_h-1)$ on the two innermost columns, whereas folding rotates the entire left and right half panels by $\pm\psi$.  The resulting main beam attenuation toward Bob outweighs the modest improvement in Eve side nulls, so the folding configuration yields a lower secrecy rate, and the bend–fold performance gap widens whenever the horizontal aperture is even.

Overall, Fig. 4 confirms that the secrecy-rate gain provided by the proposed FAA-based schemes is not limited to small proof-of-concept panels. When the array size grows from $N=9$ to $N=49$, all flexible configurations maintain a clear advantage over the rigid UPA, and the rotation-based FAA in particular exhibits an increasing rate gap as $N$ increases. This behaviour indicates that the proposed joint shape–beamforming framework scales favourably with the aperture size and can be applied to larger FAAs without incurring prohibitive computational or implementation overhead.

\subsection{The impact of an increasing number of eavesdroppers}

\begin{figure}[!t]
\centering
\includegraphics[width=3.2in]{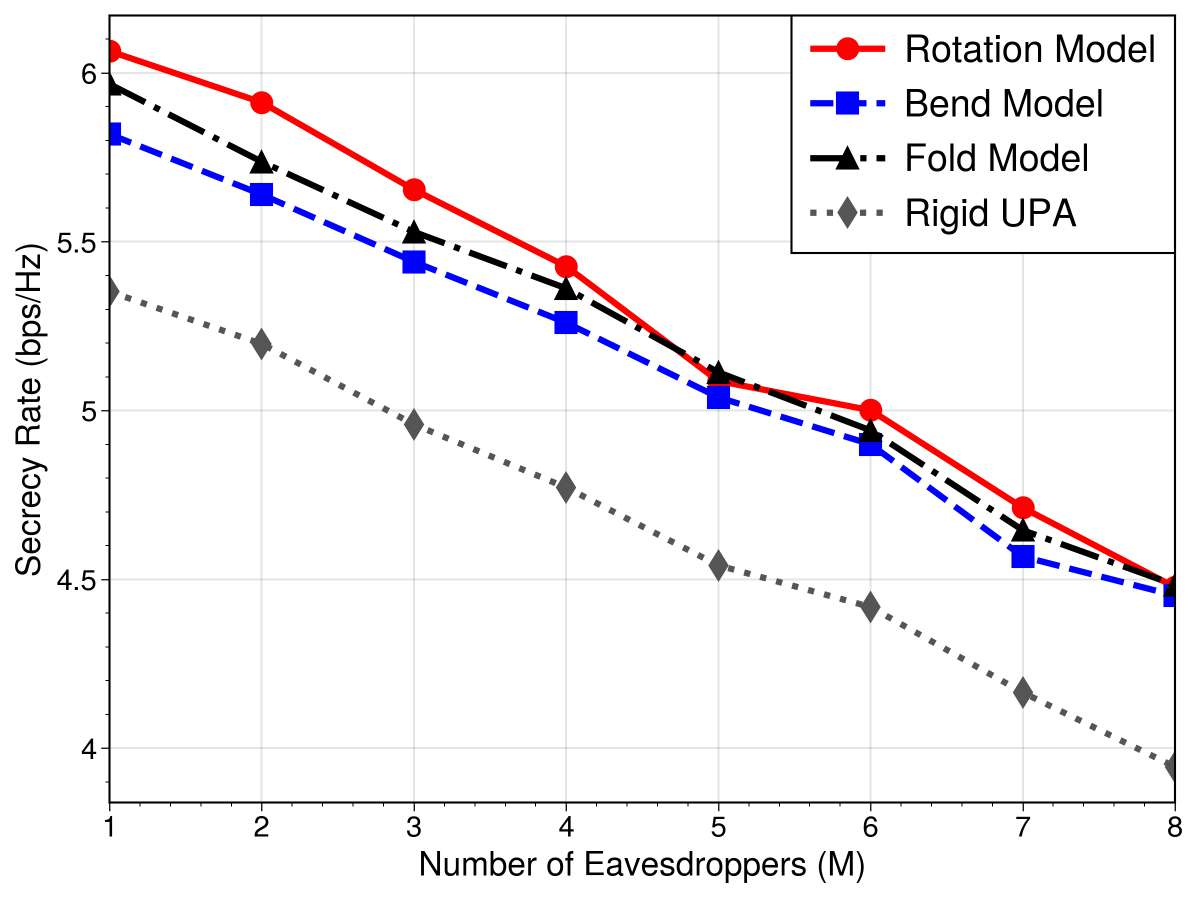}
\caption{Impact of the number of eavesdroppers on secrecy rate for different configurations.}
\label{fig_4}
\end{figure}

Fig.~\ref{fig_4} portrays the secrecy rate $R_s$ versus the number of eavesdroppers $M$ for the $3\times3$ FAA under $P_{\max}=0\;\mathrm{dBm}$ and $\kappa=4$.  By~\eqref{eq:Rs_coop_multiEve}, enlarging the eavesdroppers’ collusion monotonically erodes secrecy rate, yet every flexible array scheme consistently outperforms the rigid UPA baseline. With a single eavesdropper, the rotation model scheme achieves $6.1 \;\mathrm{bps/Hz}$. This is about $0.8 \;\mathrm{bps/Hz}$ higher than the rigid array scheme, representing an improvement of approximately 15 percent. It also outperforms the other flexible schemes by $0.2$ to $0.3\; \mathrm{bps/Hz}$. As $M$ grows, the three schemes gradually converge; At $M = 5$, they cluster near $5.0\;\mathrm{bps/Hz}$. They still maintain a margin of about $0.5\;\mathrm{bps/Hz}$ over the rigid counterpart, which corresponds to an improvement of approximately 11 percent. Even in the extreme case of $M = 8$, where the $N = 9$ transmit degrees of freedom are insufficient to null an eight-antenna eavesdropping array, the flexible shapes level off around $4.45\;\mathrm{bps/Hz}$. They still maintain a clear advantage of at least $0.55\;\mathrm{bps/Hz}$ over the rigid scheme. These results confirm that physically reshape the array—whether by rotation, bending, or folding—continues to bias the composite channel in Bob’s favour, safeguarding meaningful secrecy gains even when the eavesdroppers collusion aggressively.

\begin{figure}[!t]
\centering
\includegraphics[width=3.2in]{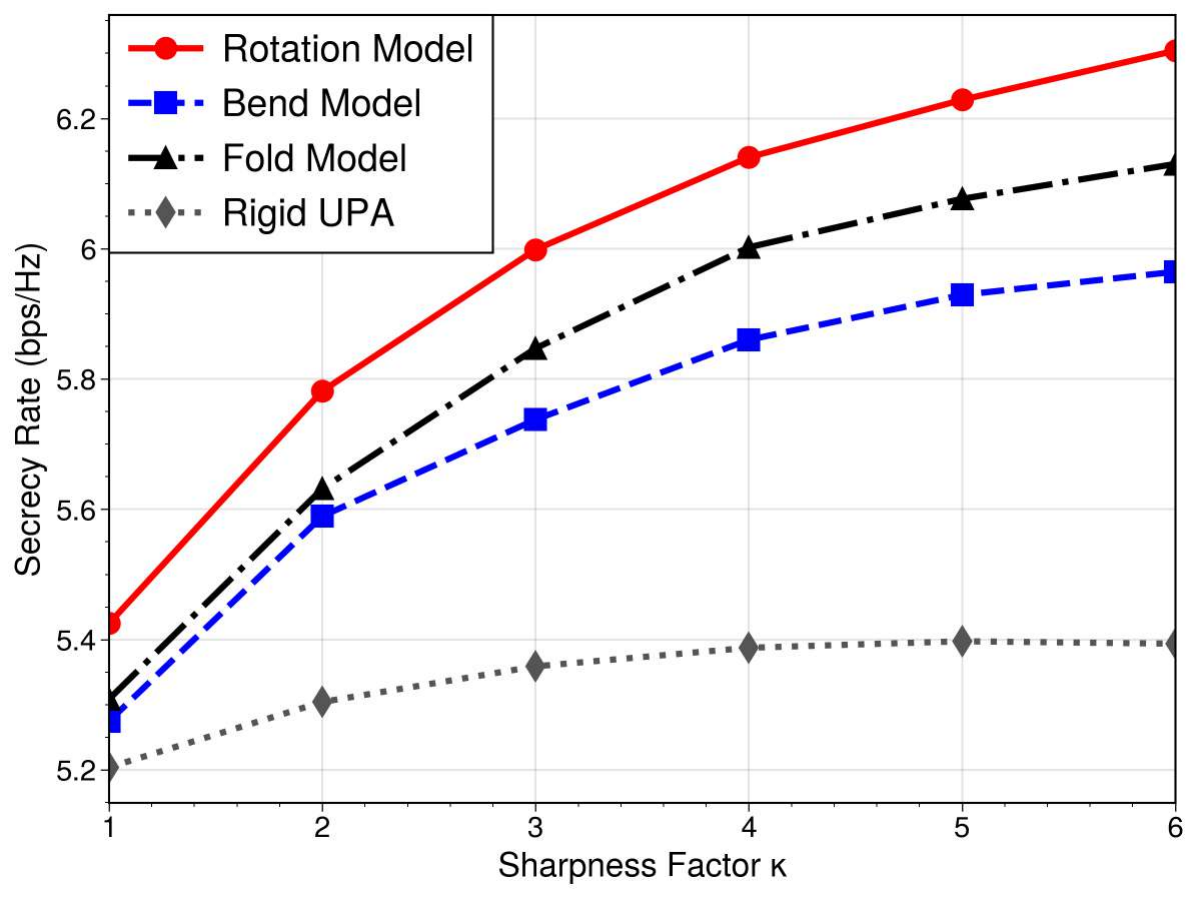}
\caption{Impact of the sharpness factor on secrecy rate for different configurations.}
\label{fig_5}
\end{figure}

\subsection{The impact of sharpness factor $\kappa$}
Fig.~\ref{fig_5} examines how the secrecy rate varies with the element–pattern sharpness factor $\kappa$ for a $3\times3$ array, single eavesdropper, and $P_{\max}=0\;\mathrm{dBm}$.  Owing to the pattern model $G_E(\theta,\phi)\!=\!2(\kappa+1)\sin^{\kappa}\theta\cos^{\kappa}\phi$, increasing $\kappa$ narrows each element’s main lobe and deepens its side‐lobe suppression, thereby reducing unintended radiation toward Eve.  Consequently, all four curves ascend monotonically; However, the rise is far steeper for the three flexible models. At $\kappa = 1$, the rotation model delivers $5.45\;\mathrm{bps/Hz}$, while the rigid UPA achieves $5.21\;\mathrm{bps/Hz}$. This corresponds to a modest gain of $0.24\;\mathrm{bps/Hz}$, or approximately 5 percent. As $\kappa$ grows to~6 the rotation curve climbs to $6.32\,\mathrm{bps/Hz}$, widening the gap over the rigid design to roughly $0.9\,\mathrm{bps/Hz}$. Bend and fold arrays follow the same trend, reaching $5.96$ and $6.15\,\mathrm{bps/Hz}$ at $\kappa=6$, respectively, and retaining $0.55$–$0.75\,\mathrm{bps/Hz}$ advantages over the rigid baseline.  In contrast, the rigid UPA saturates near $5.4\,\mathrm{bps/Hz}$ beyond $\kappa\ge4$, indicating that without geometric reconfiguration the benefit of sharper patterns quickly plateaus.  

At a more fundamental level, the behaviour in Fig.~6 can be interpreted from the large–$\kappa$ asymptotics of the element pattern in (10)–(11). For directions close to the boresight we can write $\sin\theta\cos\varphi \approx 1 - (\Delta\theta^2+\Delta\varphi^2)/2$, which yields $G_E(\theta,\varphi) \propto \exp\{-\kappa(\Delta\theta^2+\Delta\varphi^2)/2\}$. Hence each element effectively has a main-lobe width on the order of $\mathcal{O}(1/\sqrt{\kappa})$ around its boresight. For the rigid UPA, all elements share the same boresight; once Bob's dominant paths fall inside this narrow aperture while those of Eve lie outside, further increasing $\kappa$ mainly rescales $\|\mathbf{h}_B(\kappa)\|^2$ and $\|\mathbf{h}_E(\kappa)\|^2$ without creating new spatial degrees of freedom, so the secrecy rate approaches a finite ceiling, which explains the saturation of the rigid curve in Fig.~6. For the FAA models, the deformation vector $\boldsymbol{\psi}$ introduces element-wise boresight offsets, and a larger $\kappa$ amplifies the resulting per-antenna gain disparity between Bob and Eve, leading to the higher yet gradually flattening curves in Fig.~6. In the asymptotic limit $\kappa\to\infty$ the main lobe becomes so narrow that even Bob's paths eventually move outside the high-gain region; both $\|\mathbf{h}_B(\kappa)\|^2$ and $\|\mathbf{h}_E(\kappa)\|^2$ then decay towards zero and the secrecy rate $R_s$ also tends to zero. The values of $\kappa$ considered in Fig.~6 ($\kappa\leq 6$), however, correspond to practically realizable element patterns whose half power beamwidths are still on the order of several tens of degrees, so Bob remains inside the main lobe and only the saturation regime, rather than the eventual decrease, is observed in the simulated curves. The pronounced slope of the rotation curve highlights a synergistic effect: an FAA that can be rotated as a whole can align an increasingly narrow beam precisely with Bob while steering deeper nulls toward Eve, fully exploiting the added directivity.  Bend and fold models, endowed with partially independent sub‐array orientations, capture most but not all of this synergy.  Overall, the results confirm that enhancing element directivity yields the greatest secrecy dividends when combined with physical array agility.

\begin{figure}[!t]
\centering
\includegraphics[width=3.2in]{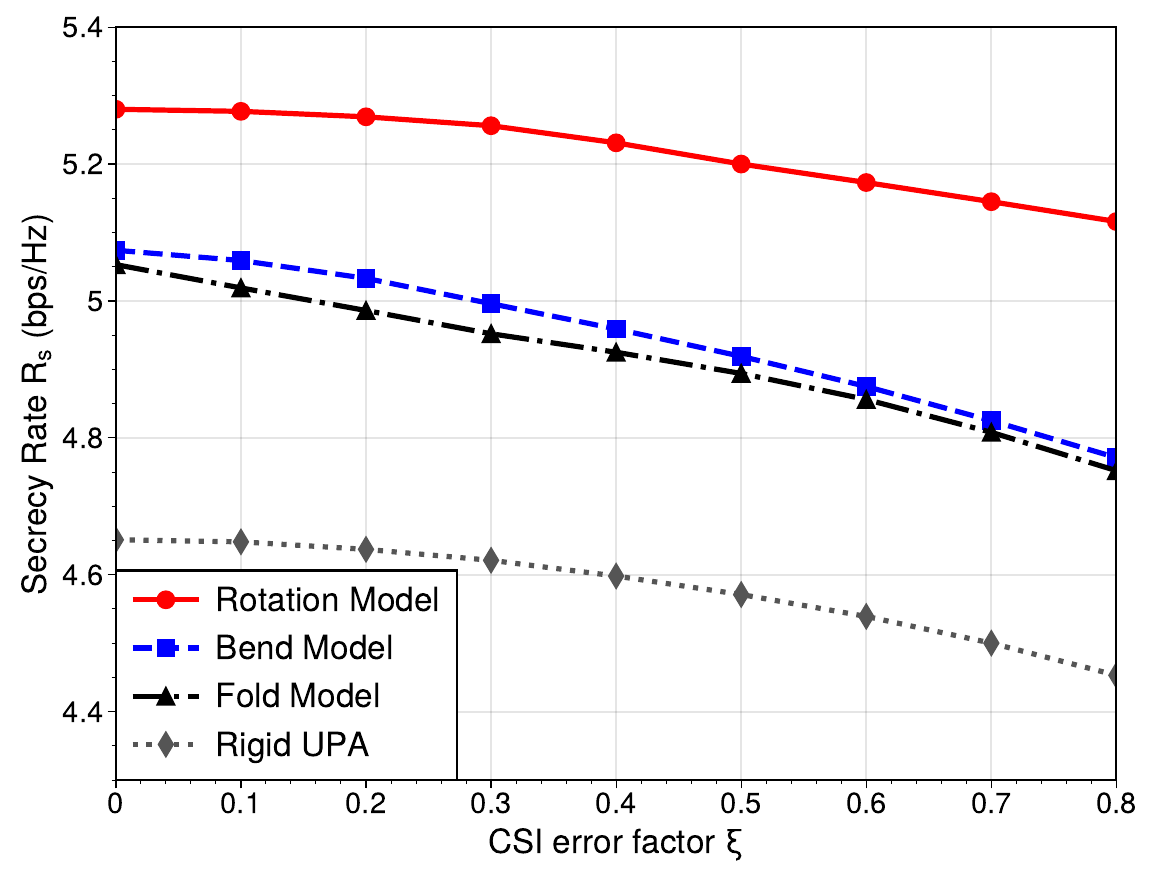}
\caption{Secrecy rate versus CSI error factor $\xi$ for different array configurations under imperfect CSI with $P_{\max}=0$ dBm.}
\label{fig_7}
\end{figure}

\subsection{The impact of CSI error factor $\xi$}
 Fig. 7 illustrates the secrecy rate versus the CSI error factor $\xi$ for the rigid UPA and the three FAA configurations when $P_{\max}=0$ dBm. As expected, the secrecy rate of all schemes decreases smoothly as $\xi$ increases from 0 to 0.8, since the estimated channels become less correlated with the true channels. The proposed rotation model always provides the highest secrecy rate, followed by the bend and fold models, while the rigid UPA has the lowest performance over the whole range of $\xi$. Even under a severe CSI error level $\xi=0.8$, the rotation-based FAA still achieves about 0.6 bps/Hz higher secrecy rate than the rigid UPA, and the relative performance loss from $\xi=0$ to $\xi=0.8$ is within only a few percent for all FAA schemes. These results confirm that the proposed joint beamforming–shape optimization is robust against moderate CSI mismatches.\\

\section{Conclusion}
In this paper, we proposed a novel secure communication system that integrates flexible antenna arrays (FAAs) with beamforming to improve secrecy performance.
By synergizing the reshape ability of FAAs with digital beam steering, our FAA‑based architecture overcomes the limitations of conventional rigid planar antenna systems.
A joint optimization framework was presented to maximize the secrecy rate through alternating optimization of the array shape parameter and the transmit beamformer.
Then, we proposed an efficient alternating optimization algorithm for the non‑convex problem, and the complexity and convergence of the algorithm were analysed.
Numerical results revealed that the proposed FAA architecture provides additional degrees of freedom for improving the secrecy rate and outperforms conventional rigid array‑based schemes. As future work, we will explore hybrid secure transmission schemes that combine FAA geometry control with other PLS designs.
Overall, the proposed FAA‑enabled scheme constitutes a promising candidate solution for physically reconfigurable secure communications.

\bibliography{Mybib}
\bibliographystyle{ieeetr} 
\end{document}